\documentstyle[12pt,aaspp]{article}
\begin{document}
\slugcomment{Revision 2.4, submitted to AJ}
\tightenlines
\def\etal{{\it et al.\/}}
\def\kms{{km~s$^{-1}$}}
\def\res{\mbox{$\cal R$}}
\title{The Luminosity Function of the Coma Cluster Core for $-25<M_R<-11$}
\author{G. M. Bernstein\altaffilmark{1}}
\affil{Dept. of Astronomy, 829 Dennison Bldg., University of
	Michigan, Ann Arbor, MI 48109 \\
	Electronic mail:  garyb@astro.lsa.umich.edu}
\author{R. C. Nichol}
\affil{Dept. of Astronomy and Astrophysics, University of Chicago, 5460 S. 
	Ellis Ave., Chicago, IL 60637 \\
	Electronic mail:  nichol@huron.uchicago.edu} 
\author{J. A. Tyson\altaffilmark{1}}
\affil{AT\&T Bell Laboratories, Room 1D432, Murray Hill, NJ 07974 \\
	Electronic mail:  tyson@physics.att.com}
\author{M. P. Ulmer\altaffilmark{1}}
\affil{ Dept. of Physics and Astronomy, Northwestern University, 
	Evanston, IL 60208 \\
	Electronic mail:  ulmer@ossenu.astro.nwu.edu}
\and
\author{D. Wittman}
\affil{Steward Observatory, University of Arizona,
 	Tucson, AZ 85721 \\
	Electronic mail:  dwittman@as.arizona.edu}
\altaffiltext{1}{Visiting Astronomer, Kitt Peak National Observatory}

\setcounter{footnote}{0}

\begin{abstract}
We determine the luminosity function (LF) of galaxies in the core of
the Coma cluster for $M_R\le-11.4$ (assuming
$H_0=75$~km~s$^{-1}$~Mpc$^{-1}$), a magnitude regime previously
explored only in the Local Group.  Objects are counted in a deep CCD
image of Coma having RMS noise of 27.7~$R$~mag~arcsec$^{-2}$.  A
correction for objects in the foreground or background of the Coma
cluster---and the uncertainty in this correction---are determined from
images of five other high-latitude fields, carefully matched to the
Coma image in both resolution and noise level.  Accurate counts of Coma
cluster members are obtained as faint as $R=25.5$, or $M_R=-9.4$.  The
LF for galaxies is well fit by a power law $dN/dL\propto L^\alpha$,
with $\alpha=-1.42\pm0.05$, over the range $-19.4<M_R<-11.4$; faintward
of this range, the galaxies are unresolved and indistinguishable from
globular clusters, but the data are consistent with an extrapolation of
the power law.  Surface brightness biases are minimized since galaxies
are not subjected to morphological selection, and the limiting
detection isophote is 27.6~$R$~mag~arcsec$^{-2}$.  We find the typical
$M_R\approx-12$ Coma cluster galaxy to have an exponential scale length
$\approx200$~pc, similar to Local Group galaxies of comparable
magnitude.  These extreme dwarf galaxies show a surface density
increasing towards the giant elliptical NGC~4874 as $r^{-1.3}$, similar
to the diffuse light and globular cluster distributions.  The
luminosity in the detected dwarf galaxies is at most a few percent of
the total diffuse light of the giant galaxies in the cluster, and the
contribution of the dwarfs to the mass of the cluster is likely
negligible as well.
\end{abstract}

\section{Introduction}

In this paper we endeavor to determine the luminosity function (LF) of
galaxies to the faintest limiting magnitudes accessible to modern
telescopes.  The LF is one of the cornerstone observations used in the
study of galaxy formation. It also has fundamental importance for the
empirical study and understanding of faint galaxy populations.  For
example, the Local Group is awash with dwarf spheroidal galaxies,
nearly all of which appear to have formed the majority of their stars
in discrete events.  At faint apparent magnitudes we might be viewing
ancestors of dwarf spheroidals undergoing star formation events, so it
behooves us to measure the density of extreme dwarf galaxies in
environments other than the Local Group.

To ease the task of identifying and counting the faintest possible
galaxies, we look where we are likely to see the most galaxies:  in the
core of the Coma cluster.  These measurements will further our
understanding of the galaxy formation process by giving direct
observations of an extreme case (low mass galaxies in dense
environments).  Luminosity functions of field galaxies can be derived
from redshift surveys of apparent-magnitude-limited samples, but such
surveys are dominated by distant giants and hence very inefficient at
finding low-luminosity galaxies.  The Stromlo-APM redshift survey of
1800 $b_J<17.15$ galaxies determines the galaxy luminosity function
only for $M_{b_J}<-15$ (Loveday \etal\ 1992); the CfA redshift survey
determines the LF for $M_{\rm Zw}\le-13$ (Marzke, Huchra, \& Geller
1994); and a recent survey by Ellis \etal\ (1995) determines the LF for
$M_{b_J}<-15$.  These surveys all find that the faint end of the LF
follows a power law $dN/dL\propto L^\alpha$ with $\alpha\approx-1.0$.

An alternative route to the galaxy LF, which we choose, is to identify
all members of a galaxy cluster, and assume that all the identified
galaxies are at a common cluster distance.  The cluster method has the
advantage that it does not require spectroscopy to obtain a redshift for
every galaxy, and therefore cluster galaxy LFs are known to fainter
magnitudes than the field LF.  The obvious disadvantage to using
clusters to determine the galaxy LF is that one measures the LF of the
cluster, not of the Universe as a whole.  Nonetheless the LF of
clusters are interesting in their own right, since differences between
cluster LFs can be direct evidence for environmental effects on galaxy
formation. In contrast, if a universal LF is derived for all clusters
it provides important insights into the underlying physics of cluster
formation.

The benchmark in the study of cluster LFs is provided by Sandage
\etal\ (1985, SBT85) with their measurement of the Virgo cluster for
$M_B<-12$. They derive a faint-end slope of $\alpha\approx-1.35$ which
is strikingly different from the value in the field. In addition to
this work, Bothun \etal\ (1991) survey both Virgo and Fornax for LSB
galaxies using photographic amplification, with a limiting surface
brightness of 27~B~mag~arcsec$^{-2}$, and found many galaxies missed by
SBT85. This demonstrates the importance of surface-brightness selection
effects in such magnitude-limited galaxy catalogs, as emphasized by
Disney (1973).  In contrast, CCDs reach lower surface brightness
thresholds as demonstrated by Turner \etal\ (1993, and references
therein) who surveyed sections of A3574 to a limiting isophote of
26.7~V~mag~arcsec$^{-2}$.  These subsequent studies have in general
agreed with the original findings of SBT85, in that the faint end of
the cluster LF follows a steeper power law than the field.

All the above studies, however, lack coverage of the field to
comparable depth as the cluster, thus giving them no rigorous means of
deciding which objects are in the cluster and which may be
foreground/background contamination.  Membership is decided primarily
on morphological grounds, which leads to a bias against galaxies which
are compact enough to resemble stars, or might be mistaken for
background galaxies; conversely, background galaxies could be mistaken
for cluster members.

We have conducted a measurement of the luminosity function of Coma
cluster galaxies which is relatively free of morphological or
surface-brightness bias, because rather than select cluster members
individually by morphological criteria, we merely count the number of
objects in an image of the Coma cluster core.  The density of
foreground and background objects is determined by counting galaxies in
images of 5 random high latitude control fields.  Simple subtraction of
the control counts from the counts in the Coma image yields a count of
excess objects, which can be assumed to be members of the Coma cluster
at redshift 7000~km~s$^{-1}$.  Because all Coma and control images are
deep CCD exposures, our magnitude and surface-brightness thresholds are
lower than in previous investigations: when optimally filtered for
objects of size 1.3\arcsec, the $3\sigma$ noise level in our images is
27.8~$R$~mag~arcsec$^{-2}$.  The linear and digital nature of the CCD
images also makes it possible to very closely match the characteristics
of the Coma and control-field images, minimizing the problem of
differential completeness which might otherwise cause error in the
subtraction of foreground/background counts from the Coma counts.  By
having 5 distinct control fields, we may also determine the
field-to-field variance on foreground/background counts, and thus can
obtain error estimates on the net Coma cluster object counts in a
model-independent manner.  Only when there is some unique aspect to the
Coma field relative to the control fields need we be concerned about
possible systematic biases in our estimated Coma cluster LF.

Use of the more distant Coma cluster rather than Virgo actually makes
the measurement easier.   If we wish to observe galaxies of a given
absolute magnitude $M$ and a certain space density in the center of a
cluster, the projected density of the galaxies on the sky will scale as
$10^{0.4(M+\mu)}$, where $\mu$ is the distance modulus to the cluster
under study.  The surface density of superposed background galaxies in
$R$ band scales almostly exactly as $10^{0.4(M+\mu)}$ as well (Tyson
1988), so that the {\it mean} ratio of target objects to background
objects is roughly independent of $\mu$.  The angular correlations of
fainter background galaxies, are smaller, however, so the {\it
fluctuations} in background counts are reduced for the more distant
cluster.  This gives better accuracy on the LF.  As long as the objects
are resolved, we pay no integration time penalty for the more distant
cluster because the time required to image the galaxy at a given S/N
increases as the square of the distance, but the solid angle subtended
by the cluster (and hence the number of separate telescope pointings
required) decreases as the square of the distance.  What is sacrificed
in a more distant survey is resolution of the target galaxies.  Indeed
in our survey we do not obtain morphological information on most of our
dwarfs.  Furthermore the faint-end limit to our LF arises not because
of S/N considerations, but because we are unable to distinguish dwarf
galaxies from globular clusters.  The resolution and sensitivity of our
images is such that most of the dwarf spheroidal galaxies known in the
Local Group would be detected if they were at the distance of the Coma
cluster, and our LF thus includes the first complete measure of the
abundance of galaxies this faint.  At $R\approx23$ ($M_R\approx-12$)
the brightest globular clusters in Coma are detected.  Harris (1987)
and Thompson \& Valdes (1987, TV87) both detect globular clusters (and
dwarf galaxies) around NGC~4874 in Coma, though in much smaller fields
than ours.  Our ability to discern dwarfs from globular clusters
disappears for $M_R>-11.4$.

We discuss in the next section the observations of the Coma and control
fields, and the production of flattened images and object catalogs for
these fields.  The image processing steps are discussed in unusual
detail, because the success of our method depends critically on the
elimination of any systematic differences in detection efficiencies
between the Coma field and the control fields; the trusting reader can
skip these details.  In \S3 we present the methodology for deriving
counts of Coma cluster members from the catalogs of Coma and control
field images.  We also enumerate several possible pitfalls in the
application of this methodology to our data, concluding that none of
them are of serious concern.  In \S4 the methodology is applied to
derive the luminosity and size distributions of the members of the Coma
cluster core, and the variation of their surface density with distance
from the cluster center.  In \S5 we look in closer detail at the LF of
galaxies in the Coma cluster, comparing our results to those of other
clusters by other authors.  We also estimate the masses required of
these objects for tidal integrity.  In \S6 we discuss several possible
evolutionary scenarios for the dwarf galaxies in Coma, and their
relation to dwarfs in the field, and we give a brief summary and
suggestions for future observations in \S7.  An extensive review of the
properties of dwarf elliptical (dE) galaxies and their evolution is
given by Ferguson \& Binggeli (1994, FB94).   Because the breadth of
knowledge and speculation on dwarf galaxies is so large, we refer the
interested reader to this review and minimize our rehashing of the
literature. Though we have little direct knowledge of the morphology of
the $M_R\sim-12$ galaxies in the Coma cluster, observations of nearer
clusters detect few dwarf irregular galaxies, and these gas-rich
galaxies are unlikely to be present in the core of Coma.  We will
therefore, for simplicity, often refer to the dwarfs in Coma as dE
galaxies. We will assume a Hubble parameter of
$H_0=75$~km~s$^{-1}$~Mpc$^{-1}$, giving a distance modulus of 34.9 for
the Coma cluster.  Under this assumption, 1\arcsec\ subtends 460~pc at
the Coma cluster, and our field spans a 200~kpc square.

\section{Creating the Object Catalogs}

\subsection{Observations}

The Coma field was observed on 10~February~1991 from the KPNO 4-meter
telescope, using a backside-illuminated $1024\times1024$ Tektronix CCD
at prime focus.  One pixel spans 0.473\arcsec\ on the sky, giving an
8\arcmin\ field of view.  A series of $27\times 300$~second exposures
was taken in a ``shift and stare'' mode (Tyson 1986):  the telescope
pointing is different for each exposure, as much as 1\arcmin\ from the
nominal field center.  After de-biasing and flat-fielding
(discussed below), the relative
offsets of the exposures are determined to $\sim0.1$~pixel by
photometry of bright stars, and the images are registered using
bilinear interpolation.  Note that the interpolation introduces
correlations between adjacent pixels, so that the noise spectra of our
images are not white.  At each pixel, the signals from all 27~exposures
are averaged, after a $3\sigma$ clip to remove cosmic ray events.  The
shifting reduces the high-S/N area of the final image to a
7.5\arcmin\ square.  The FWHM of the PSF in the final $R$-band image is
1.3\arcsec.  This image is shown in Figure~\ref{rawpic}.  The field is
centered at approximately $12^{\rm h}57^{\rm m}30^{\rm s}$, $+28\deg
09\arcmin 30\arcsec$ (1950), near the x-ray centroid of the cluster
(Ulmer, Wirth, \& Kowalski 1992).  The giant elliptical galaxies
NGC~4874 and NGC~4889 lie 40\arcsec\ and 280\arcsec\ off the NW and NE
corners, respectively, of our frame.  Exposures in the $B_j$ filter
were also obtained, but the seeing was too poor for these data to be
useful for investigating the faintest Coma cluster galaxies.

The flat field for this run is produced in a two-step process:  the
small-scale structure is determined from dome flats.  The large scale
structure of the flat is determined from the median of 44 disregistered
300~second exposures of 7 distinct high-latitude fields which are free
of large objects.  These exposures were all taken during the same run
as the Coma images, and are distinct from the high-latitude data used
to determine background counts.  This median blank-sky flat is then
divided by the dome flat and smoothed, and the resulting image is
multiplied by the dome flat to give the final flat-field image.  When
``blank-sky'' exposures are processed using this flat-field, the
remaining variations in the sky level are much smaller than the
observed diffuse-light gradients in the Coma field, so we conclude
that the measured diffuse signal is not significantly affected by
flat-fielding errors.

Five regions of the sky were observed under nearly identical conditions
for use as control fields in this study.  These fields are selected at
random, subject to the constraints that they have low extinction in the
Burstein \& Heiles (1982) map, and be free of bright ($R\lesssim14$)
stars or galaxies.  The characteristics of these control fields are
listed in Table~1.  Although these control exposures were not taken
during the same run as the Coma image, they are all in the same filter
band, all are taken at prime focus of either the KPNO or CTIO 4-meter
telescopes, and all using thinned Tek1024 CCDs.  The observing,
flat-fielding, and image-combining techniques are the same as for the
Coma field.

All of the fields were observed through the Kitt Peak Harris $R$
filter, or the CTIO version of same.  Two of the control fields contain
faint standard stars from Tyson \& Seitzer (1988, TS88); for the
remainder, photometric zeropoints were transferred from observations of
a TS88 field on the same night.  The TS88 $R$ band is for these
purposes identical to the Cousins $R$ system.  The Coma observations of
February~1991 were non-photometric, with up to 0.7~mag of obscuration
by clouds.  The Coma field and a TS88 standard field were re-observed
with the same setup on the photometric night of 14~June~1991 to
determine the photometric zeropoint of the deep Coma image.  We
estimate the zeropoint accuracy of the fields to be 0.05~mag or better,
based on variance of standards and a few repeat measurements.

\subsection{Matching of Control Fields}

The success of our differential counting method depends upon matching
the detection characteristics of the control fields {\it as closely as
possible} to those of the Coma field.  Most of the control field images
have better seeing and/or lower noise than the Coma image, so we
degrade them to better match Coma.  We first degrade the seeing of a
control image by convolving it with a Gaussian of proper width to bring
the FWHM of the PSF near the Coma FWHM of 2.77~pixels. The FWHM of the
processed control fields, listed in Table~1, range from 2.71 to
2.81~pixels.  Smoothing the control images in this way removes
high-frequency noise---we therefore next add high-pass-filtered noise
back into the images in order to restore the original noise power
spectrum.

Three of the control images have lower noise than Coma, so we add noise
to these images.  The RMS sky signal fluctuations in the Coma image are
29.32~mag per pixel (surface brightness 27.7~mag~arcsec$^{-2}$); the
control fields have noise levels from 29.19 to 29.37~mag per pixel.
The noise power spectra, not just the RMS levels, are matched.  In
Table~1 we list the noise spectral densities of the Coma and matched
control-field images.  These numbers give the amplitude of the flat
part of the noise power spectrum in our images; recall that some
short-wavelength noise is removed by the bilinear interpolation during
image registration.  To estimate the RMS noise when the images are
smoothed by a rectangular window of area $A$, multiply by $1/\sqrt{A}$;
if the smoothing window is a Gaussian with dispersion $\sigma$,
multiply by $1/\sqrt{\pi \sigma^2}$.  It is important to specify the
resolution or window size when quantifying a surface-brightness noise.

The control fields are thus very similar to the Coma field in
resolution and noise, and hence in detection efficiency.  We conduct
Monte Carlo completeness measurements as described below to quantify
the small remaining differences in detection efficiency among the
fields.  When reading the following sections, keep in mind that in
\S\ref{radial} we detect a very strong gradient of Coma cluster members
toward NGC~4874.  If our signals were merely errors in magnitude scale,
extinction corrections, or completeness corrections, we would not
expect to find such a gradient.

\subsection{Removal of Diffuse Light}

\label{diffuse}

We cannot search for faint objects using the image shown in
Figure~\ref{rawpic} because the diffuse light gradient overwhelms the
detection algorithms.  We must remove the large-scale gradients from
the Coma image in order to successfully use the FOCAS programs.  We
describe here the procedure used for this; because this procedure was
not executed for the control fields, we must also be sure that it does
not alter the signals from small objects in the Coma field, lest the
subtraction of control counts be invalidated.  The first step in the
removal of the large-scale features of the image is to fit elliptical
isophotes to the 23 brightest galaxies (and one bright star).  The
fitted ellipses are then subtracted from the image, allowing FOCAS to
successfully search for dwarf galaxies where it previously was
``blinded'' by flux gradients due to giant galaxies.  Regions where the
elliptical isophotes are a poor fit are masked and ignored in further
processing.  Once these larger objects are subtracted, we fit the
large-scale gradients in the image by running a 15\arcsec-square median
filter across the image.  These steps of galaxy fitting and
diffuse-light subtraction must be iterated for best results,
particularly in the NW corner of the image, which contains steep
diffuse-light gradients from NGC~4874, another bright S0 galaxy, and a
bright star.

The final step in the removal of the diffuse light is to use the FOCAS
program {\sc detect} to create a diffuse-light map of the field, and
then subtract away this diffuse component.  FOCAS tracks the sky level
during its detection phase using an exponential average with a scale
length of $\approx40\arcsec$.  This subtraction of the FOCAS ``sky''
image is done for the control fields as well as for the Coma field.
The resulting cleaned and masked Coma image is shown in
Figure~\ref{cleanpic}, and the many low-luminosity objects are now
clearly visible.

In Figure~\ref{rawpic} it appears that the diffuse light gradient from
NGC~4874 extends all the way across our image (see also the diffuse
light plot in Figure~\ref{radfig}).  Is this diffuse light really
present in the Coma cluster, or are we merely seeing scattered light
from the core of NGC~4874 ($R\approx11$) or the $m_V\approx7$ stars
SAO~082595 and SAO~082589 located within 30\arcmin?  The tabulation of
the distant wings of the typical seeing disk by King (1971) suggests
that the scattered light should be much weaker than the observed
diffuse light.  This is verified by our own tests with the CTIO 4m
prime focus camera, in which we image ``blank'' high-latitude fields
within 1\deg\ of bright stars and search for diffuse scattered light.
Since the CTIO and KPNO 4m telescopes are very similar, we conclude
that our diffuse-light signal originates in the Coma cluster, and is
not an artifact of atmospheric or telescope optics.

\subsection{Catalogs}

The FOCAS software package (Valdes 1989) is used to detect and analyze
objects in the six fields.  For the uninitiated, the properties of
FOCAS of relevance to this work are as follows:  all objects are found
which exceed local sky by more than $3\sigma$ (27.6~R~mag~arcsec$^{-2}$
in our images) in a filtered version of the image.  On the unfiltered
images these objects are checked for multiple maxima by an
object-splitting algorithm.  Several photometric quantities are
calculated for each object using the unfiltered image.  The two we use
here are the ``total'' magnitude $m_t$, and the ``core'' magnitude
$m_c$.  The former is derived by growing the object several pixels in
all directions around the original detection isophote and measuring the
flux above sky inside this expanded region.  This quantity $m_t$ is
shown by Monte Carlo tests to be an unbiased estimator of the true
magnitude of unresolved objects.  Our tests show that it remains
unbiased for Gaussian-profile objects of any size, but does
underestimate total magnitude for exponential-profile objects which are
near the detection limit and have scale sizes large compared to the
seeing disk (we will return to this question in \S\ref{sizesec}).  The
core magnitude is the highest flux found in any contiguous $3\times3$
pixel subarea of the image, and thus is roughly the peak surface
brightness.  We will use ${\cal R}\equiv m_t-m_c$ as our resolution
parameter, since unresolved objects have a nearly fixed value of \res.
We prefer \res\ to the FOCAS ``scale'' parameter because the former
takes continuous values and is more robust in crowded areas.

FOCAS is run with the same parameters for all six fields, except that
the detection threshold is maintained at fixed increment above the
(slightly varying) noise level.  The detection threshold is chosen so
that the number of false positive detections is negligible---Monte
Carlo tests suggest that there are $\sim15$ detections of noise with
$25<R<25.5$, and none with $R<25$.

The final step in producing object lists is to delete from the catalogs
those objects centered within 5\arcsec\ of any edge or masked area of
the image.  The useful area $A_i$ of each field, excluding the masked
areas and 5\arcsec\ buffer zones, is listed in Table~1.

\subsection{Exclusion of Bright Foreground Stars}

\label{stars}

In Figure~\ref{resfig} we plot the $R$ magnitude (the $m_t$ values
produced by FOCAS) versus the resolution parameter \res\ for all of the
valid objects in the Coma field.  At bright magnitudes the stellar
locus at constant $\res\approx0.65$ is apparent.  For $R<21$ we may
confidently identify objects with $\res<0.8$ as stellar; at fainter
magnitudes the galaxy population begins to merge into the stellar
locus, so we do not attempt object-by-object classification.  The
$R$-vs-\res\ plot for the 5 control fields are similar, and we identify
as stellar those objects with $R<21$ and $\res<0.8$.

\subsection{Extinction Correction}

\label{extinct}

Counts of extragalactic background objects will be depressed by
Galactic dust extinction, so we must compensate for this extinction if
we are to use the control fields as models for the Coma background
counts.  We use the maps of Burstein \& Heiles (1982) to determine the
extinction in each field relative to that of Coma, assuming an $R$-band
extinction $A_R=2.5E_{B-V}$.  $A_R$ is only 0.03~mag in the Coma field
itself.  The detected $R$ magnitudes of all objects in each control
fields are adjusted brightward by the relative extinction corrections
listed in Table~1. These corrections bring the control fields into
agreement with the Coma extinction, not to zero extinction.  The
correction of 0.12~mag in the Ser1 field is the largest, and most of
the other fields have insignificant extinction.

Henceforth we will use only these extinction-corrected $R$ magnitudes.

\subsection{Completeness Tests}

\label{monte}

Completeness estimation is straightforward because of the automated
nature of the cataloging process.  If we are interested in the
detection efficiency $f_i$ in image $i$ for objects of a given
magnitude and shape, we simply add objects with these properties into
image $i$ and run FOCAS exactly the same way as for the original
catalogs.  We then search the new catalog for detections within
1.5\arcsec\ of the locations of the artificial objects; $f_i$ is then
simply the fraction of the artificial objects which are recovered.
Artificial objects which merge with brighter preexisting objects are
not considered to be ``recovered.'' These Monte Carlo tests give us the
{\it marginal} detection probability $f_i$ for objects with the chosen
magnitude and shape---the probability that the next additional object
would be detected.

We have conducted these Monte Carlo tests to obtain the completeness
$f_i$ vs.\ magnitude for stellar objects in each of the fields, with
results shown in Figure~\ref{compfig}.  The Coma image is 50\% complete
for stars at $R=25.5$, and the control fields match this value to
within 0.1~mag.  Because the fields are well-matched and the $f_i$ are
very similar, the completeness corrections have only a second-order
effect on the removal of background counts from Coma.  At $R\sim24$ the
Coma field is less complete than the control fields because it contains
many more large galaxies which obscure parts of the field.  The very
brightest of these we have either masked or subtracted away using
elliptical isophote fits, but many $R>16$ cluster members remain, and
effectively ``shadow'' the $R>22$ galaxies in which we are primarily
interested.  The Monte Carlo results for $f_i$ are necessary to
quantify this effective loss of area in the Coma field due to increased
crowding.  Note that the ``bright'' galaxies which cause the crowding
have $M_R\lesssim-13$, well into the regime of what are normally
denoted as dwarf galaxies.  Our chief interest lies in yet fainter
galaxies.

The Monte Carlo tests show that $m_t$ is an unbiased estimator of true
magnitude for stellar objects down to the 50\% completeness level.  The
RMS error in magnitude for $R=25.25$ stellar objects is 0.4~mag, with
of course smaller errors at brighter flux levels.

We have also conducted extensive Monte Carlo tests using resolved
objects instead of stellar objects.  The techniques used are exactly
the same as above; we will defer discussion of these results until
later, when we investigate the sizes of Coma cluster galaxies.

\section{Methodology of Differential Counts}

\subsection{Estimating Background Counts}

We wish to use data from the 5 control fields to estimate the number of
objects in the Coma field which are background or foreground to the
Coma cluster itself.  For brevity, we will henceforth use the word
``background'' to mean objects that are either foreground or background
to the Coma cluster.  We assume that the number of and detection
efficiency for the background in the Coma field is {\it typical} of the
control fields.  In other words, the Coma {\it cluster} objects have
simply been added into our catalogs atop the background typical of high
latitude fields.  If this is true, we may estimate the background
counts in the Coma field by taking the mean of the control field
counts.  We may furthermore estimate the {\it uncertainty} in the Coma
background counts from the variance of the control-field counts.  In
the next subsection we discuss ways in which the Coma field might have
an atypical background population or perhaps an atypical detection
efficiency for background objects.  First we give the formulae used to
estimate the background and its uncertainty.

Consider a particular type of object---{\it e.g.} having detected
magnitude within a certain range.  We wish to estimate the number
$N_{bg}$ of these in the Coma catalog which are background as opposed
to cluster members.  Let the number of equivalent objects in the
control fields be $N_1,N_2,\ldots,N_M$, where $M=5$ is the number of
control fields.  Since $N_{bg}$ is assumed to be drawn from the same
population as the $N_i$, our best estimate for $N_{bg}$ is simply the
mean
\begin{equation}
\label{mean}
\tilde N_{bg} = {1 \over M} \sum_{i=1}^M N_i { {A_C f_C} \over {A_if_i} }.
\end{equation}
We make corrections (of a few percent) to the counts $N_i$ to correct
the areas $A_i$ and detection efficiencies $f_i$ of the control fields
to match $A_C$ and $f_C$ of the Coma field.  Next we ask what will be
the RMS deviation $\sigma_{bg}$ of the actual $N_{bg}$ from the
estimate $\tilde N_{bg}$.  The best estimate of this is given by
\begin{equation}
\label{var}
\sigma_{bg}^2 = { {M+1} \over {M(M-1)}} 
		\sum_{i=1}^M (N_i { {A_C f_C} \over {A_if_i}} -\tilde N_{bg})^2.
\end{equation}
In essence we have drawn $M$ samples from some distribution, and ask
how accurate our estimate of the next sample will be.  Note that this
is not the same as estimating the variance of the population, as is
usually done, and hence we have an extra factor of $(M+1)/M$ relative
to the usual expression.

We can use Equations~(\ref{mean}) and (\ref{var}) to estimate the
background counts in some magnitude bin, or in some joint
magnitude-\res\ bin.  The beauty of this technique is that error
estimates are generated from the control fields, so we need not rely on
models of our uncertainties.  Background fluctuations due to Poisson
statistics, galaxy clustering, magnitude zeropoint errors,
mis-estimation of the $f_i$, etc., are all automatically included in
$\sigma_{bg}$ as long as the processes which generate these variations
are operating similarly in the Coma field as in the control field.
Thus we do not have to estimate or even be aware of the sources of
error in the background correction, as long as these errors are the
same in the Coma field as in the control fields.

\subsection{Estimating Cluster Counts}

If $N_C$ objects are detected in the Coma field and $\tilde N_{bg}$ is
our estimate of the number of background objects, then the remainder
\begin{equation}
\label{ndet}
N_{det} = N_C - \tilde N_{bg}
\end{equation}
is our best estimate of the number of {\it detected} cluster members.
The uncertainty $\sigma_{det}$ in $N_{det}$ is equal to $\sigma_{bg}$,
since there is no uncertainty in $N_C$.  Next we wish to estimate the
number $N_{cl}$ of objects actually present in the Coma cluster within
the solid angle of our field.  We must correct for the detection
efficiency $f_C$, so
\begin{equation}
\label{ncl}
N_{cl} = N_{det} / f_C.
\end{equation}
Finally we desire the uncertainty in this quantity, for purposes of
fitting luminosity functions, etc.  We assume that galaxies are placed
into our field by a Poisson process (a cosmic variance of sorts), and
that the detection process is also Poissonian.  Then the final
uncertainty in the parent cluster luminosity function is given by
\begin{equation}
\label{sigcl}
\sigma^2_{cl} = (\sigma^2_{bg} + N_{det}) / f_C^2,
\end{equation}
where the first term is the background uncertainty and the second term
accounts for the Poisson fluctuations of cluster counts.  Note that the
fluctuations in the background counts are already contained within
$\sigma^2_{bg}$.

\subsection{Possible Pitfalls}

Having many control fields allows us to make internal estimates of both
the background and its variance for our program field, free of
systematic biases, as long as the program field is not extreme in some
way relative to the controls.  There are, however, several ways in
which the Coma field is extreme relative to the controls, and we must
assure that these differences do not bias our results.  Here we
enumerate these potential pitfalls and their effects.

\subsubsection{North Galactic Pole---Dust and Stars}

The Coma field would be unique even without the presence of the Coma
cluster because it lies near the North Galactic Pole, and thus suffers
less Galactic dust extinction than the average control field, and is
likely to have lower star counts as well.  We have corrected for
extinction as outlined in \S\ref{extinct}, which should alleviate the
first problem.  Differential star counts should not be a problem for
$R<21$ because we have excluded stars from the object counts as
described in \S\ref{stars}  For $R>21$, stars are not readily
distinguished from galaxies, so we do not attempt an exclusion of
stellar objects because we do not want to be biased against compact
galaxies in the Coma cluster.  We expect, though, that the difference
in star counts between the NGP and our control fields is small compared
to our cluster signals.  We have calculated the expected NGP star
counts using the Galactic models of Ratnatunga (1993), and these are
plotted in the lower panel of Figure~\ref{compfig} along with the total
object counts.  If the Coma field were to have fewer stars than the
control fields, this would cause us to {\it underestimate} the Coma
cluster galaxy counts in the $21<R<23$ range.

\subsubsection{Dust in the Cluster}

The Coma field differs from the control fields in one other major
respect, namely that there is a large galaxy cluster in the middle of
it!  The presence of the Coma cluster could alter the appearance of
galaxies in its background in two ways.  First, if there were dust in
the cluster core, the background galaxy counts would be diminished.
Romani \& Maoz (1992) detect a deficit of quasars in Abell cluster
fields, and posit an extinction of $E_{B-V}\sim0.3$ in cluster cores as
an explanation.  Ferguson (1993), however, limits $E_{B-V}<0.05$ in the
Coma cluster based on the colors of elliptical galaxies.  We assume
zero extinction in the cluster.  If there is dust in Coma, it reduces
the number of observed background galaxies and again causes us to {\it
underestimate} the number of cluster member galaxies.

\subsubsection{Gravitational Lensing}

The Coma cluster might also alter the appearance of its background
population by gravitational lensing. The critical surface mass density
in Coma for lensing of distant background galaxies is
$\Sigma_c=c^2/4\pi GD = 3.5$~g~cm$^{-2}$.  A singular isothermal sphere
with a 1-d velocity dispersion of 1000~km~s$^{-1}$ has a surface mass
density of $\Sigma=0.24 (r/100\,{\rm kpc})^{-1}$~g~cm$^{-2}$, where $r$
is the distance from the core; if the cluster has a finite core radius
$r_c$, the central surface mass density is obtained by setting
$r=r_c$.  Distant background objects will be magnified by a factor
$(1-\Sigma/\Sigma_c)^{-1}$ in area, which of course increases the
flux from each background galaxy (gravitational lensing conserves
surface brightness).  The cluster lens amplification, however, {\it
reduces} the solid angle of distant Universe being viewed behind Coma
by the same factor.  If the background object counts per magnitude
scale as $d(\log N)/dm=\alpha$, then the lensing will change the
background counts by the factor
$|1-\Sigma/\Sigma_c|^{2.5(0.4-\alpha)}$.  In the $R$ band,
$\alpha=0.39$ (Tyson 1988), so we see that lensing will have a very
small effect upon the background counts (in fact will {\it decrease}
the counts) unless the surface density of Coma is near the critical
value.  To reach $\Sigma=\Sigma_c$ would require a core radius for Coma
of $r_c\le7$~kpc, or 15\arcsec\ at the distance of Coma.  Thus even in
the very unlikely event that the Coma cluster is a critical lens, the
critical region would subtend only a small part of our field, and
furthermore the lensing would cause us to again {\it underestimate} the
Coma cluster membership.  We may safely ignore the effects of
gravitational lensing!

\subsubsection{Large-Scale Structure}

One may worry that our estimate of the Coma cluster LF may be
artificially inflated because the Coma field crosses the Great Wall, a
large sheet-like concentration of galaxies in which the Coma cluster is
embedded.  While structures such as the Great Wall are not rare in the
Universe, and hence may also be present in the control fields, the
reader may be concerned that we are in fact measuring the LF of the
Coma core {\it plus} a cross section of the Great Wall population.  In
\S\ref{radial}, we find that the Coma field excess objects are strongly
clustered on NGC~4874, indicating that the majority of these excess
objects are probably true cluster members rather than superposed Great
Wall members.

\subsubsection{Diffuse Light Gradient}

The Coma image has a very strong diffuse light gradient which is not
present in any of the control field images.  This might have two
effects on the detection of background objects:  first, the diffuse
light contributes photon noise and surface-brightness-fluctuation noise
to the Coma image which is absent from the controls.  This could
potentially affect the detection efficiency or false positive rate for
faint objects.  Calculations indicate that this effect is negligible,
and this is borne out by the fact that artificial galaxies placed in
the higher-noise part of the Coma image are detected with the same
efficiency as identical galaxies placed in the lower-noise half.

The second effect of the diffuse light is more worrisome:  it has
forced us to perform an image processing step on the Coma image which
was not performed on the control images, as described in
\S\ref{diffuse}.  The galaxy-fitting and median-filtering steps remove
only long-wavelength signals from the image, and hence have little
effect upon objects with angular sizes below 10\arcsec.  The control
catalogs contain few or no faint background galaxies this large, so
there is no bias in the Coma background galaxy estimate.  It is
possible, however, that the median filter has removed flux from
extremely diffuse Coma cluster member galaxies, and therefore {\it
depressed} the apparent number of Coma cluster members in our final
LF.  Our Coma catalog does become incomplete for faint LSB objects with
scales $\gtrsim10\arcsec\approx5$~kpc because they are attenuated by
the median filtering and FOCAS sky-subtraction; a few such objects are
visible in the unfiltered Coma image.  Proper study of these most
diffuse objects in the Coma field requires a different approach to
background removal, and we defer discussion of these objects to a later
paper (Ulmer \etal\ 1995). Their numbers are small, however, and will
not affect the conclusions of this paper.

\subsubsection{Crowding}

The Coma image has of course a much higher density of bright galaxies
than any of the controls, leading to a lower faint-object detection
efficiency for Coma because of the effective loss of area.  As already
discussed, we compensate for this crowding by first subtracting or
masking the very brightest galaxies, and then running Monte Carlo tests
to calibrate the ``shadowing'' effect of the remaining galaxies.
Crowding affects the detected counts in a non-linear fashion, and hence
when the density of faint objects in the Coma field greatly exceeds
that in the control fields, it may be inaccurate to estimate the
crowding effects through the addition of a few Monte Carlo galaxies at
a time to the control images.  A complete Monte Carlo simulation
requires that we add to the control images an artificial version of the
{\it entire} population of galaxies posited to exist in the Coma
cluster, and note whether the Coma counts are recreated.  We have
implemented a few such full-population Monte Carlo tests, and the
results are in agreement with those derived from the simpler,
model-independent technique of estimating detection efficiencies by
adding a few stars at a time to each image.  We conclude that the
simpler technique adequately accounts for crowding.

\section{Magnitude, Size, and Spatial Distributions of Coma Cluster
Members}

In this section we apply the methodology of \S3 to the catalogs
described in \S2, deriving the luminosity function and typical sizes of
galaxies in the Coma cluster.  We also investigate the spatial
distribution of the very faint cluster members.

\subsection{Magnitude Distribution}

We may apply Equations~(\ref{mean}) through (\ref{sigcl}) to the
numbers $N_i$ of objects with a given range of magnitudes to obtain an
estimate of the number $N_{cl}$ of cluster members in that magnitude
bin.  In Figure~\ref{compfig} we plot the Coma counts $N_C$ vs. $R$
magnitude, alongside the estimate and uncertainty $\tilde N_{bg}$ and
$\sigma_{bg}$ of the background counts derived from the control field.
Also shown in Figure~\ref{compfig} are the expected numbers of
background galaxies and Galactic stars.  The Coma counts significantly
exceed the control-field counts at almost all magnitudes down to our
completeness limit, as seen in Figure~\ref{compfig}.

In applying Equations~(\ref{mean}) through (\ref{sigcl}) to the data in
a given magnitude bin we use for $f_i$ the detection efficiency derived
for {\it unresolved} objects at that magnitude.  Coma cluster members
could be resolved and have somewhat lower detection efficiencies; hence
our derived values of $N_{cl}$ may be low estimates of the true
population near the limiting magnitude of the image.  The 50\%
completeness magnitude for exponential-profile galaxies of various
scale lengths is plotted in Figure~\ref{sizefig}.

In Table 2 we give the gross counts $N_C$, background $\tilde N_{bg}$,
net $N_{det}$, and completeness-corrected counts $N_{cl}$ in each
magnitude bin.  Figure~\ref{lffig} plots as filled circles the derived
$N_{cl}$ vs.\ $R$ magnitude, the Coma cluster luminosity function
(LF).  Parameterizing the LF as a power law $dN/dL\propto L^\alpha$,
two regimes are apparent:  for $R<23.5$ ($M_R<-12.4$), the LF is
consistent with the $\alpha\approx-1.3$ range that the deepest previous
cluster LF studies have found (Sandage, Binggeli, \& Tammann 1985,
SBT).  For $R>23$ ($M_R>-12.9$) we find a much steeper LF, with
$\alpha\approx-2$. At these magnitudes, however, it is possible that
Coma cluster globular clusters are being detected as well as Coma
cluster galaxies.  We will therefore defer further discussion of the
Coma cluster galaxy LF until the next section.  We first investigate
the characteristic sizes and the spatial distribution of the detected
Coma cluster members, which will help in distinguishing galaxies from
globular clusters.

\subsection{Extension to Bright Galaxies}

The brightest galaxies in the Coma field are saturated on our CCD
images, and furthermore there are few bright galaxies in our small
field.  In order to extend our LF measurement to brighter limits, we
have used the catalog of Godwin, Metcalfe, \& Peach (1983, GMP83), who
have digitized $b$- and $r$-band 2.5\deg\ square photographic images of
the Coma cluster.  Their published galaxy catalog is complete to
$b=21$, and thus we can use it for galaxies $r<18$ without danger of
incompleteness.  For the 21 unsaturated galaxies in our catalog with
$R<19$, a fit for a magnitude offset between our and the GMP83
photometry yields $R = r+0.29\pm0.17$.  We use this to convert the
GMP83 $r$ magnitudes into our $R$ system.  The surface density of
galaxies differs across the face of the Coma cluster, and we want to
generate a bright-galaxy LF which is directly comparable to our CCD
field, so we must sample the GMP83 data at comparable distances from
the Coma center as our CCD data.  We generate a LF from the GMP83
sample using only the galaxies within 8\arcmin\ of NGC~4874.  The CCD
field consists essentially of one quadrant of this circle.  As a
background region for the GMP83 8\arcmin\ target sample we use an
annulus at $1\deg<r<1.25\deg$ from NGC~4874.  The density of galaxies
in this annulus is $\le$20\% of the density inside the
8\arcmin\ circle, so we need not be too concerned about whether this
background annulus contains cluster members.  We subtract the
background galaxy density from the target galaxy density and multiply
by the area of the CCD field to obtain bright-galaxy counts to compare
with the CCD counts. Uncertainties are derived from the square root of
the target-field counts.  The GMP83 galaxy counts are plotted as the
open circles in Figure~\ref{lffig}, and are listed in Table~3.  The
agreement between the CCD data and the GMP83 data is good in the region
of overlap.  In this way we obtain a LF for Coma cluster members for
the range $10<R<25.5$, spanning more than 6 decades in luminosity!

\subsection{Are the Faint Objects Resolved?}

\label{resolved}

Previous observations of faint objects near NGC~4874 have found
globular clusters as bright as $B\sim24$ (TV87; Harris 1987), and have
yielded hints of a population of resolved objects as well.  Recent HST
observations may have observed globular clusters as bright as $V=24.2$
near NGC~4881, another Coma cluster giant elliptical galaxy (Baum
\etal\ 1994).  This suggests that the excess objects in our Coma
field may include globular clusters at $R\gtrsim23.5$ as well as dwarf
galaxies.  Our PSF FWHM of 1.3\arcsec\ subtends 600~pc at the distance
of Coma, so the globular clusters would be unresolved in our images,
while the dwarf galaxies may or may not be resolved.  In order to
determine the possible extent of globular cluster contamination in our
LF, and to determine the size range of our dwarf galaxies, we
investigate the distribution of the resolution parameter \res\ for our
excess galaxies.

When determining the Coma cluster LF, we did not actually attempt to
assign cluster membership to individual objects, but rather used a
statistical method to determine the overall number of cluster members.
Likewise, for $R>21$, it is impossible to unambiguously determine the
degree of resolution of each individual object.  This is because the
distribution of \res\ for unresolved objects begins to overlap the
\res\ distribution for galaxian objects.  We can, however, determine
the degree of resolution for the entire population by comparing the
observed distribution of \res\ to the distribution expected for a
population of unresolved objects.

In Figure~\ref{slicefig} we plot the distribution of \res\ for the Coma
cluster objects in each of six magnitude slices.  These histograms were
derived in exactly the same way as the luminosity function:  we count,
for example, the number $N_C$ of objects in our Coma catalog having
$23.5<R<24.0$ {\it and} $1.0<\res<1.25$.  We likewise count the numbers
$N_i$ of objects in this range of $R$ and \res\ in each of the control
catalogs.  The methods of \S3 are then used to derive the number
$N_{det}$ and uncertainty $\sigma_{det}$ of detected excess objects in
this joint magnitude-resolution bin.  This process is repeated over a
two-dimensional grid of $R$-\res\ bins, with the results shown in
Figure~\ref{slicefig} as the histograms with error bars.

Also shown in Figure~\ref{slicefig} as the dotted curves are the
distributions in \res\ that would be expected if all the excess counts
were unresolved.  The \res\ distributions for unresolved objects are
obtained from the Monte Carlo simulations.  The PSF template used for
the Monte Carlo unresolved objects is a Moffat function, which has
$I(r)\propto(1 + r^2/r_0^2)^{-\beta}$.  The two parameters $\beta$ and
$r_0$ are adjusted to force both the FWHM and the second-moment radius
to match those of bright stars in the Coma image.

It is immediately apparent from the first two panels of
Figure~\ref{slicefig} that in the $22.5<R<23.0$ and $23.0<R<23.5$
magnitude ranges {\it most of the Coma excess consists of resolved
objects.}  Indeed we can make a conservative estimate of the number of
galaxies in these magnitude bins by counting the objects with $\res>1$,
since we do not expect any unresolved objects to be detected with such
large \res\ values.  Our estimates of the number $N_{res}$ of resolved
objects are shown in Table~2, and are plotted as the open triangles in
Figure~\ref{lffig}. The great majority of objects in these two bins are
seen to be resolved by this simple test, or by other, more
sophisticated, tests.  The same holds true in all brighter magnitude
bins, which are not plotted.

Note that the uncertainties in the number of definitively resolved
objects ($\res>1$) are smaller than the error bars on the total number
of objects.  This is true because most of the variance in background
counts at these magnitude ranges is for unresolved or marginally
resolved objects ($\res<1$).  This is partially due to the fact that
cluster dwarf galaxies tend to be more diffuse than background
galaxies, a fact well known to the many astronomers (such as SBT85) who
have sought dwarf galaxies in more nearby clusters using morphological
criteria.  The agreement apparent in Figure~\ref{lffig} between our
resolved counts (open triangles) and the overall counts (filled
circles) is reassuring evidence that there are not large numbers of
compact dwarf galaxies in the Coma cluster for $M_R\lesssim-12$.  This
is good news if true in Virgo and Fornax as well, because it means that
there are not large numbers of {\it high} surface brightness dwarfs
that SBT85 and Ferguson \& Sandage (1988, FS88) would have missed in
their morphologically selected cluster LF's.

In the next fainter bin, $23.5<R<24.0$, there is a tendency for the
detected Coma cluster objects to have \res\ greater than expected of
unresolved sources, but the data are consistent with an unresolved
population at a one-sigma level so we cannot reject this hypothesis.
Likewise we find that the \res\ histograms in all of the fainter bins
are consistent with a population of unresolved objects.

We cannot conclude, however, that all of the excess objects with
$R>23.5$ are globular clusters.  As we near our completeness limit of
$R=25.5$, several factors conspire to make it difficult to distinguish
between galaxies and globular clusters:  first, the detected galaxies
at $R<24$ are clearly becoming smaller as their luminosities
decrease---note that in the first three panels of
Figure~\ref{slicefig}, the peak of the \res\ histogram is moving toward
the left (less resolved).  If we extrapolate this apparent
size-luminosity relation, we find that for $R>24$, we {\it expect} most
of the dwarf galaxies to be too small to resolve.

At fainter magnitudes, the S/N of our detections is getting too low to
permit much discrimination as to their size.  Monte Carlo tests show
that for $R>24.5$ it becomes essentially impossible to distinguish a
population of stellar objects from a population of objects 1.5--2 times
as large as the PSF.  Objects significantly more extended become
seriously incomplete at these magnitudes, as they fall below the
surface brightness threshold of our detection process.  Thus for
$R>24.5$ the \res\ histogram will {\it always} be indistinguishable
from that of an unresolved population due to S/N limitations.

The most direct evidence that many of the excess $R<25$ objects must be
galaxies is given by TV87, who have imaged at 0.6\arcsec\ FWHM
resolution a field located 40\arcsec\ W of NGC~4874.  They use a field
340\arcsec\ away from NGC~4874 as a ``reference'' field, though in fact
this second field is within our larger Coma image\footnote{As a
cautionary note, we point out that the ``prominent dwarf galaxy with
many subcondensations'' within the TV87 reference field is completely
absent in our deeper image.  Ghost images are easily confused with very
extended galaxies.}.  In the magnitude range $25.25<B<26.25$,
corresponding roughly to $24<R<25$, TV87 report approximately 30 more
objects in the inner field than in the reference field.  Of these,
roughly one-third are resolved and are denoted by TV87 as a ``retinue
of dwarf galaxies'' near NGC~4874.  The remainder are unresolved and
are assumed to be globular clusters.  This suggests that a substantial
fraction of the $23.5<R<25$ objects in our field are in fact galaxies
which we cannot resolve.

To summarize, examination of the distribution of the resolution
parameter \res\ for the excess objects gives unambiguous evidence that
we may consider the great majority of the $R<23.5$ ($M_R<-11.4$) excess
counts to be galaxies.  For $R>23.5$, we cannot identify a resolved
population, but external evidence suggests that a substantial fraction
are dwarf galaxies while globular clusters could easily comprise the
majority.

\subsection{Size Distribution of Coma Cluster Galaxies}

\label{sizesec}

We may use the observed \res\ distributions to determine a
characteristic size for the Coma cluster dwarf galaxies as a function
of magnitude.  As a means of quantifying the sizes of these dwarfs, we
will assume that they have exponential surface-brightness profiles,
\begin{equation}
\label{expdisk}
I(r) = I_0 e^{-r/r_s},
\end{equation}
and circular symmetry.  Of course we have little information on the
profiles of the extreme dwarf galaxies in our images.  The best-studied
galaxies of similar luminosities are the Local Group dwarf spheroidals,
which are well described by exponential profiles, so for ease of
comparison we adopt this function.  Indeed almost all faint dE galaxies
are satisfactorily fit by exponentials (FB94; Impey, Bothun \& Malin
1988).

We perform Monte Carlo tests to determine the distributions of detected
magnitude and \res\ to be expected if galaxies of a given luminosity
and $r_s$ are present in the Coma cluster.  The IRAF task {\sc artdata}
is used to add artificial seeing-convolved exponential galaxies onto
the Coma image, and the methods described in \S\ref{monte} are used to
determine the detection efficiency and expected parameter
distributions.  We can then see if exponential-profile galaxies with
the selected magnitude and $r_s$ are present in Coma by noting whether
this Monte Carlo distribution of $R$ and \res\ overlaps significantly
with the Coma R--\res\ distribution in Figure~\ref{slicefig}.

The results of these comparisons are shown in Figure~\ref{sizefig}.  We
restrict our attention to the faintest objects in our CCD frame.  For
the $22.5<R<23.0$ bin, corresponding to $M_R=-12.2$, we find that the
Coma galaxies' scale lengths are in the range $250\,{\rm
pc}<r_s<450\,{\rm pc}$.  Objects 0.5~mag fainter are about a factor 2
smaller.  This range of detected galaxy sizes is shown as the heavily
shaded region in Figure~\ref{sizefig}.  For galaxies fainter than
$R\approx23.3$ ($M_R>-11.6$), our detections are not meaningfully
resolved.  The lightly shaded region in Figure~\ref{sizefig} denotes
the magnitudes and scale sizes that could potentially exist in the Coma
cluster as part of our unresolved detections---this region extends to
zero size, of course.  Finally, the heavy line shows the locus at which
our detection efficiency is 50\%.  Galaxies rightward of this line
would likely have escaped detection by us.  In producing
Figure~\ref{sizefig} we have made corrections for the tendency for
FOCAS to underestimate the flux from exponential-profile objects, so
that the magnitude scales refer to actual rather than detected
magnitude.

Only in the Local Group is comparable information available for
galaxies this faint.  The 10 Local Group dwarf spheroidal galaxies for
which scale lengths and magnitudes are available are plotted on
Figure~\ref{sizefig} as they would appear at the distance of the Coma
cluster (the Fornax galaxy is too bright to appear on this plot).
These data are primarily from Caldwell \etal\ (1992), and we have
assumed $V-R=0.5$ for the dSph's.  We note first that 7 of these 10
galaxies would have been detected in our images if they were placed in
the Coma cluster.  Second, we note that {\it the sizes of the Coma
cluster dwarf galaxies are similar to Local Group dSph's of comparable
magnitude.}

The observed size distribution of Coma cluster members does not abut
the detection threshold for $R<24$.  In other words, we have found a
minimum to the central surface brightness of galaxies.  We
are thus unlikely to be missing significant numbers of $R<24$ galaxies
due to surface brightness selection effects, unless of course there is
an entirely disjoint population of objects lurking beneath our surface
brightness threshold of $\approx27.6$~mag~arcsec$^{-2}$.

\subsection{Radial Distribution of Coma Cluster Members}

\label{radial}

Information on the spatial distribution of different types of cluster
members could be important in determining their natures and histories.
Examination of Figure~\ref{cleanpic} suggests that the object density
in Coma increases markedly toward NGC~4874, and indeed a measurement of
the object density as a function of the distance $r$ from the center of
NGC~4874 shows a significant gradient for all types of objects.  We bin
the detected objects in the Coma field into several annuli around
NGC~4874, and calculate the fraction $a$ of the entire useful field
that lies within each annulus.  The background estimates $\tilde
N_{bg}$ from the control fields are scaled by $a$ to give an estimate
of the background in each annulus, and we scale $\sigma_{bg}$ by $\sqrt
a$ to estimate the uncertainty in the background within each annulus.
We then calculate the object density above background in each annulus
of the Coma image.  These excesses are shown as a function of $r$ for
the magnitude range $20.5<R<22.5$ as squares in Figure~\ref{radfig}.
The starred symbols in Figure~\ref{radfig} show the radial gradient for our
$22.5<R<23.5$ objects; these are the faintest objects which we can
reliably label as galaxies.  The open circles in Figure~\ref{radfig}
show the radial gradient of the $23.5<R<25.5$ Coma objects, which may
be predominantly globular clusters.  In each magnitude bin we actually
plot the mean surface brightness above background contributed by Coma
cluster members.  No incompleteness corrections have been made, so the
surface brightness in the faintest bin may be underestimated by a few
tenths of a magnitude.

Also plotted in Figure~\ref{radfig} is the strength of the Coma cluster
diffuse light signal visible in Figure~\ref{rawpic}.  The hatched area
reflects the uncertainty in our knowledge of the background diffuse
brightness level.  At one extreme we have assumed that the SE corner of
our image, 10\arcmin\ from NGC~4874, contains {\it zero} diffuse
cluster light.  The upper bound of the hatched region shows the diffuse
light signal obtained by assuming that the $r^{-1.3}$ power-law
gradient in diffuse light seen in the NW corner of our image extends to
$r=10\arcmin$.  The dotted lines in Figure~\ref{radfig} all trace
$r^{-1.3}$ power-law profiles, for reference.

Finally we have included in Figure~\ref{radfig} the radial profiles of
the light provided by galaxies in the GMP83 catalog, divided into
$14<b<17.5$ and $17.5<b<21$ magnitude bins.  The error bars on these
points also arise from our uncertainty in the background galaxy
density.  We must derive the background density from the
$0.9\deg<r<1.1\deg$ annulus of the GMP83 catalog.  The error bars on
the plotted points are the bounds obtained by assuming that between
10\% and 90\% of the counts in this outer annulus are due to background
galaxies.

Several conclusions may be drawn immediately by inspection of
Figure~\ref{radfig}.  Foremost, we see that the excess counts we detect
in Coma show a highly significant gradient toward NGC~4874 at all
magnitudes.  This is very strong evidence that our excess counts are in
fact due to objects present in the Coma cluster.  If our excess counts
were an artifact of some background subtraction error ({\it e.g.\/} one
of the pitfalls mentioned in \S3), then we would {\it not} expect to
see this gradient.  Similarly, if the excess were due to Great Wall
objects not physically associated with the Coma cluster, we would not
expect a concentration about NGC~4874.

Our next immediate conclusion is that the dwarf galaxies and globular
clusters contain little of the total luminosity of the cluster.  Within
the $1\arcmin<r<10\arcmin$ region spanned by our CCD field, we see that
the total diffuse flux is similar to the flux from bright galaxies in
the cluster.  What would normally be called ``dwarf'' galaxies, with
$17.5<b<21$, contribute $\approx1.5$~mag less light at
$r\gtrsim5\arcmin$, while the extreme dwarfs from our CCD image
($R>20.5$) and globular clusters produce 6--7~mag less light than the
diffuse light or giant galaxies.  Unless these extreme dwarfs have
mass-to-light ratios several hundred times larger than the giant
galaxies, they do {\it not} contain a major fraction of the cluster
mass.

We note also that the radial density gradient differs for various
classes of objects.  Parameterizing the surface brightness as
proportional to $r^\beta$, we find that the diffuse light is best fit
by $\beta=-1.3\pm0.1$.  The same $\beta=-1.3$ power law is a good fit
to both magnitude slices of the GMP83 catalog for $r>5\arcmin$, but the
``dwarf'' galaxies at $17.5<b<21$ show a significant flattening of the
density for $r<5\arcmin$.  This conclusion was also reached by Thompson
\& Gregory (1993), who report a drop in density of dwarf spheroidal
galaxies for $r<20\arcmin$, and a somewhat milder flattening for their
``very faint dE'' sample at $b\approx20.5$.  We have not subdivided our
objects by morphology, but there seems to be general agreement.
Thompson \& Gregory suggest that their dSph galaxies are being
destroyed by the cluster tidal field.  It is thus interesting to see if
the central deficit persists to yet fainter galaxies.  In fact we find
that the squares in Figure~\ref{radfig}, representing our $20.5<R<22.5$
dwarfs, follow a $r^{-1.3}$ power law for $r>1.5\arcmin$, but are
significantly depleted for $r<1.5\arcmin$; this turnover radius is
quite a bit smaller than the $r\approx5\arcmin$ at which the GMP83
galaxy density (filled triangles) flattens out.  Note that our results
for the radial gradients of dwarf galaxies are quite similar to the
results of Vader \& Sandage (1991), who find a dwarf density $\propto
r^{-1.22}$ in the vicinity of RSA elliptical galaxies.  Their data show
a deficit of dwarfs relative to this power law for projected radii less
than $50h^{-1}$kpc, which would correspond to 2.4\arcmin\ at the
distance of the Coma cluster.  The Vader \& Sandage dwarfs are selected
on a morphological basis, but would correspond roughly to a 17--20 $R$
magnitude range if placed in the Coma cluster.  Zaritsky \etal\ (1993)
find 69 satellites of 45 spiral field galaxies, and find the projected
surface density of satellites drops as $r^{-1.0\pm0.2}$ within 300~kpc
of the primaries.  This is consistent with our population about
NGC~4874.

The faintest objects in our images follow a $\beta\approx-1.3$ density
gradient over our entire field.  The starred symbols 
in Figure~\ref{radfig} show
the run of density with $r$ for the faintest objects which can
confidently be called galaxies, $22.5<R<23.5$.  The open circles, for
$23.5<R<25.5$ objects, also are close to a $\beta=-1.3$ line, so that
these fainter objects---be they $M_R>-11$ galaxies or globular
clusters---share the steep gradient of the diffuse light.  If tidal
destruction is the cause of the central hole in the Thompson \& Gregory
dwarf spheroidal density or the central plateau in the $17.5<b<21$
galaxy density, then the fainter galaxies we detect must be more robust
to tidal stress.  The plateau region becomes smaller for fainter
galaxies, and is entirely absent for $M_R>-12$ galaxies in our data.
We will consider the tidal stability of our objects in the next
section.

Any concentration of objects toward the x-ray centroid of the cluster
is at least 10 times weaker ($2\sigma$ upper limit) than the
concentration toward NGC~4874.  Indeed the x-ray emission itself peaks
on NGC~4874, though the x-ray isophotes are quite asymmetric, being
stretched toward NGC~4889.  The x-ray brightness gradient within
several arcmin of NGC~4874 is extremely weak, with $\beta\approx-0.15$
(Dow \& White 1995).  The strong peak in faint galaxy density and
diffuse red light towards NGC~4874 thus lies inside a much flatter
x-ray gas distribution.

We have searched for any tendency for the density of faint Coma cluster
members to increase near the locations of bright galaxies other than
NGC~4874.  In particular we ask whether there is any concentration of
objects toward the 10 brightest E/S0 galaxies in our field (excluding
NGC~4874).  For distances of 24\arcsec\ to 120\arcsec, we find no
change in density of $22<R<25.5$ objects.  Our sensitivity is such that
a 20\% change in the density of Coma cluster members would be noticed.
Within 24\arcsec\ of the E/S0 galaxies we note a $2\sigma$ decrease in
faint galaxy density; given the possible completeness problems this
close to a bright galaxy and the poor statistics, we do not consider
this significant.  It thus appears that the number of globular clusters
or dwarf galaxies which are bound to the generic giant galaxies in the
cluster is small compared to the number bound to the potential well
surrounding NGC~4874.  An interesting exception is the tight subcluster
of galaxies visible on the left side near the lower edge of
Figures~(\ref{rawpic}) and (\ref{cleanpic}).  The bright S0 galaxy in
this subcluster is indeed a cluster member with $cz=7895$~km~s$^{-1}$
(Mazure \etal\ 1988); it would be worthwhile to obtain redshifts for
the other galaxies in this apparent clump, which are redder than
the central S0.

\section{The Galaxy Population of the Coma Cluster Core}

In this section we take a closer look at what the data presented above
tell us about the galaxies found in the core of the Coma cluster,
particularly with reference to results on very faint galaxies in other
environments.  We will work primarily with absolute magnitudes and
sizes rather than apparent units.  As a reminder, if we parameterize
the Hubble constant as $H_0=75h_{75}$~km~s$^{-1}$~Mpc$^{-1}$, we are
assuming $h_{75}=1$.  For other Hubble constants, luminosities scale as
$h_{75}^{-2}$ and sizes scale as $h_{75}^{-1}$.

\subsection{Luminosity Function for Galaxies $M_R<-11.4$}

Our first task is to determine the slope of the faint end of the galaxy
LF.  We fit power laws of the form $dN/dL\propto L^\alpha$ to the
$N_{cl}$ data in Tables 2 and 3 with $-19.4<M_R<-11.4$.  The bright
limit of the fit is chosen to be a few magnitudes below the canonical
$L^\ast$ for clusters ({\it e.g.\/} Lugger 1986), placing us on the
power-law part of the LF.  This also sidesteps the vagaries of
small-number statistics at the very bright end.  We are fitting only
toward the faint side of the possible ``hole'' in the Coma cluster LF
at $b=17.5$ ($M_R\approx-19$) reported by Biviano \etal\ (1995).  We do
not attempt to fit faintward of $R=23.5$ since globular cluster
contamination may be a problem there.  We use the resolved-count
estimates $N_{res}$ for $21<R<23.5$ rather than the total counts
$N_{cl}$.  While this introduces a morphological bias against compact
dwarf galaxies, it reduces the uncertainties.  If we instead fit to the
$N_{cl}$ we get results that are completely consistent, albeit with
slightly larger errors.  Likewise we obtain consistent results whether
or not we include the photographic data at the bright end.

Least-squares fitting over this range yields $\alpha=-1.42\pm0.05$,
with a $\chi^2$ value of 7.2 for 9 degrees of freedom.  This power law,
shown as the solid line in Figure~\ref{lffig}, is thus completely
consistent with the data.  The 95\% confidence interval is
$-1.57<\alpha<-1.25$.  These $\alpha$ values are consistent with those
derived for Coma by Biviano \etal\ (1995) and by Thompson \& Gregory
(1993), who each conduct much shallower ($M_R\lesssim-16.5$) but
wider-area surveys of this cluster.

This $\alpha$ is compatible with most results on the faint-end slopes
of cluster LFs, or of dE galaxies in clusters.  An overview of previous
results is in FB94; here we update this discussion in light of our
results.  The most extensive previous measurement of the LF of a galaxy
cluster is that of SBT85, who have selected Virgo cluster members on a
morphological basis to a limit of $B=18$.  Assuming a Coma/Virgo
distance ratio of 5.5, and a typical $B-R$ color of 1.3 for dwarf
galaxies, this limit corresponds to $M_R=-14.4$ with our assumed
$H_0$.  To this depth, SBT85 derive an overall LF with $\alpha=-1.25$
(no uncertainties given).  A slightly steeper slope of $\alpha=-1.30$
is fit to incompleteness-corrected counts to $B=20$ ($M_R=-12.4$).  A
fit to our Coma counts over the $-19.4<M_R<-12.4$ region yields
$\alpha=-1.32\pm0.07$, with $\chi^2=3.32$ for 7 degrees of freedom.
Thus our Coma LF has a very similar shape to the SBT85 Virgo LF at the
faint end.  Various morphological subsets of the SBT85 data are fit
with faint-end slopes of $-1.45<\alpha<-1.35$; we have not made
morphological distinctions, but still measure similar slopes at very
faint magnitudes.  FS88 measure the Fornax LF to similar depth as the
SBT85 Virgo study---the overall LF slope in Fornax is consistent with
the Virgo and our Coma results.  Several caveats are in order, however,
before we happily conclude that all clusters have $\alpha\approx-1.4$
at the faintest measured magnitudes.

\subsubsection{Very Steep Cluster Luminosity Functions?}

De~Propris \etal\ (1995) have recently reported very steep faint-end
LFs in several rich Abell clusters.  In particular, they fit
$\alpha=-2.2\pm0.2$ to A2199 galaxies with $-15.5<M_B<-10.5$,
corresponding for typical dwarf colors to $-16.8<M_R<-11.8$.  This
differs by many sigma from our results for similar galaxies in Coma.
De~Propris \etal\ report similarly steep slopes for the I-band LF of
three other clusters.  Such dramatic differences between the Virgo and
Coma clusters on the one hand, and A2199 on the other, deserve further
investigation.  These authors lack sufficient control field coverage
and were forced to use galaxy counts from the literature to correct for
background.  It would take a rather large error in background
correction to bring their LF slope into agreement with ours, but counts
of background galaxies are notorious for varying when measured by
different investigators.  These authors have repeated their
measurements with more extensive background observations and we await
their results.

Driver \etal\ (1994) measure the $R$-band LF of the $z=0.206$ cluster
A963, to a limit of $R\approx24.5$, which corresponds to
$M_R\approx-16$.  The overall slope of the A963 LF over the
$-21<M_R<-16$ range is $\alpha\approx-1.5$, not greatly different from
our lower-redshift Coma data.  These authors prefer to interpret their
data as the sum of two populations, one with $\alpha=-1.0$ and another
with $\alpha=-1.8$, but no other data are available in this cluster to
support a distinction into two populations.

\subsubsection{Varying Dwarf/Giant Ratios}

If different galaxy types have distinct faint-end LF slopes, then the
faint-end slopes of overall LFs in different environments could vary
due to changes in the mix of galaxy types.  In cases where the
morphologies of the dwarf galaxies are known, one may fit LFs to the
distinct types.  SBT85 fit Schechter functions to various types of
Virgo dwarfs, obtaining faint-end slopes in the range $-1.35$ to
$-1.45$.  FS88 conclude that while the overall LF in Fornax resembles
that in Virgo, the dE population in Fornax has a significantly
shallower slope ($\alpha=-1.09\pm0.09$) than in Virgo.  Ferguson \&
Sandage (1991) further investigate the dwarf population in various
nearby environments, and posit a general rule that the early-type
dwarf-to-giant ratio increases significantly with cluster richness.
While we do not have morphological types against which to test this
relation, Ferguson \& Sandage (1991) offer another formulation of this
hypothesis, which states that the ratio of faint ($-17<M_R<-15$) galaxy
counts to bright ($M_R<-18$) counts also increases with cluster
richness.  Our data indicate that this ratio is lower in our portion of
the Coma cluster than in the Virgo cluster, which would counter the
posited relation.  A more effective demonstration of this is given by
Thompson \& Gregory (1993), who have morphological information for
nearly the entirety of the Coma cluster galaxies with
$M_R\lesssim-16.5$.  They conclude that the early-type dwarf-to-giant
ratio in Coma is no higher than in Virgo.

\subsubsection{Radial Variation of Luminosity Function}

Another caveat to the direct comparison of Figure~\ref{lffig} with the
SBT85 and similar data is that we image only the central portion of the
Coma cluster.  If the LF varies with radius from the cluster center,
then our cluster core LF differs from the LF of the cluster as a
whole.  Thompson \& Gregory (1993), for example, suggest that certain
types of diffuse dwarf galaxies are deficient near the center of Coma.
Figure~\ref{radfig} also suggests that galaxies with luminosities in
the LMC range ($-17.4<M_b<-14$) may be less concentrated on NGC~4874
than either brighter or fainter galaxies, which roughly follow a
$r^{-1.3}$ density gradient over the range in which we measure them.
In the Virgo cluster, the faint galaxies seem to be well mixed with the
giant galaxies (FB94).  Would we measure a different $\alpha$ if we
were to survey the entire Coma cluster for $M_R<-11.4$ dwarfs?
Examination of Figure~\ref{radfig} suggests that the LF for the entire
cluster would, if different, be shallower than our derived
$\alpha=-1.42$ were we to include more peripheral regions of the Coma
cluster.

We check for radial variation of the LF by splitting our CCD field into
two equal-area regions, one with $r<5.4\arcmin$ and the other
$r>5.4\arcmin$.  In each half we count the number of resolved objects
($\res>1$) in bins spanning $15.5<R<23.5$.  We scale the background
counts by one half and the background uncertainties by $1/\sqrt{2}$,
correct for incompleteness, and fit to power-law LF's as for the full
sample.  We find that for the inner half of the field ($r<5.4\arcmin$),
$\alpha=-1.50\pm0.08$, with a 95\% confidence interval of
$-2.0<\alpha<-1.2$.  For the outer half of the field,
$\alpha=-1.25\pm0.11$, with a 95\% confidence interval of
$-1.6<\alpha<-0.5$.  The fitted LF is shallower at greater distances
from NGC~4874, but this is a weak signal.  In fact both halves of the
field are consistent with the $\alpha=-1.42$ determined from the entire
sample.  We do not detect any significant change in LF across our
field, but the test is rather weak.  The recent observations by Secker
\& Harris (1994) should, when fully reduced, give a more definitive
answer to the question of radial dependencies in the LF in Coma.

\subsubsection{Surface Brightness Biases}

There is growing evidence that surface brightness selection effects
have an important impact on measurements of the LF in various
environments.  Ferguson \& McGaugh (1995) demonstrate how a single
galaxy population can potentially be responsible for the variety of
measured $\alpha$ values in the field, as various surveys have distinct
surface brightness thresholds.  Sprayberry \etal\ (1995) construct a
field LF from a galaxy survey with low surface brightness threshold,
and likewise find that inclusion of LSB galaxies changes the LF
significantly.  Measurements of cluster galaxy LFs are not immune to
surface brightness selection effects.  We show here, however, that our
Coma LF measurement is, for the reasons mentioned in the introduction,
much less susceptible to surface brightness biases than previous
works.

Each survey has a bounded region of sensitivity in the surface
brightness vs.\ absolute magnitude ($\mu, M$) plane.  In deriving their
Virgo LF, SBT85 corrected for insensitivity to LSB objects by noting a
correlation of surface brightness with absolute magnitude in the
complete region of this plane, and extrapolating into the incomplete
region.  Impey, Bothun, \& Malin (1988) conducted a further survey of
Virgo dwarfs using photographic amplification, and uncovered 26 new
potential LSB Virgo dwarfs, most occupying regions of the ($\mu, M$)
plane inaccessible to SBT85.  They repeat the SBT85 extrapolation, this
time to lower central surface brightness of $\mu_0\approx26$ in $B$,
and suggest that $\alpha=-1.7$ in a range comparable to our
$-17<M_R<-13.5$ (no error cited).  After a further study of Fornax
dwarfs, Bothun, Impey, \& Malin (1991) attempt extrapolation into
unexplored regions of the ($\mu, M$) plane using a slightly different
method:  they assume that the LF is a separable function of scale
length $r_s$ and central surface brightness $\mu_0$, and derive
$\alpha=-1.6\pm0.2$.  This result is consistent with our Coma result
(and with the SBT85 result, formally), but it does inspire us to assess
our SB biases.

Two kinds of SB bias are possible:  firstly against {\it high} SB
galaxies which are missed in morphologically-based surveys due to their
resemblance to foreground stars or background giant galaxies.  We are
free of such biases because we (initially) imposed no size restrictions
on our counts.  As previously discussed, our results do not change
significantly when we restrict ourselves to resolved objects, and we
thus conclude that there are few high SB dwarfs in the Coma cluster.

The survey is inevitably biased against {\it low} SB galaxies which are
so extended as to fall below the SB threshold of the detection
process.  It is difficult to define a single SB threshold for our
images, because the detection process for very extended objects is not
limited by photon noise, but rather by the gradients in the diffuse
cluster light.  We do, however, have a surface brightness threshold
well below any of the Virgo or Fornax studies---the detection isophote
is nominally 27.6~$R$~mag~arcsec$^{-2}$, whereas the limiting isophote
for the Impey, Bothun, \& Malin (1988) study is
$\mu\lesssim27$~$B$~mag~arcsec$^{-2}$.  In a more rigorous test of our
sensitivity to LSB objects, we place artificial versions of
exponential-profile galaxies on the Coma image and see if they survive
the sky subtraction and detection process.  Most of the galaxies
detected via photographic amplification in Fornax have
$r_s\lesssim10\arcsec$ and $B\lesssim20$; these galaxies would appear
as $r_s\lesssim2\arcsec$, $R\lesssim22.7$ in our Coma data.  Monte
Carlo tests show that we would detect at least 75\% of them, although
the diffuse-light subtraction and FOCAS local sky determination lead to
as much as 1.5~mag underestimate of their luminosity.  The
incompleteness is primarily due to crowding rather than noise.
Likewise Impey, Bothun, \& Malin (1988) find in Virgo a few galaxies
with scale lengths as large as 30\arcsec, at $B\lesssim17$.  Such
objects, if present in the Coma cluster, would be immediately visible
in Figure~2, and would be efficiently detected, albeit with significant
lost light.

We thus believe that we are capable of detecting galaxies at much lower
central SB than in previous photographic surveys of clusters.  We do
{\it not} find the galaxy population filling the ($\mu, M$) plane to
the faint-$\mu$ limit of our detection region, except for the faintest
galaxies as depicted in Figure~\ref{sizefig}.  In other words, there
does seem to be a locus of maximum galaxy density in this ($\mu, M$)
plane, and it is well within our detection region for $M_R<-12$.  The
existence and tightness of a correlation between $\mu$ and $M$ for dE
galaxies is under current debate (see FB94), and our extremely deep CCD
images with wide dynamic range may help settle this question, at least
as regards the Coma cluster.  We defer further discussion of this
question, however, to the succeeding paper (Ulmer \etal\ 1995), because
the analysis of the most extended galaxies in our image requires us to
remove diffuse light in a different manner.

We conclude that our LF is less subject to SB biases than any previous
effort because the image has such low noise.  While our algorithms
underestimate the luminosities of the most diffuse objects, we find few
objects large enough to be significantly affected in this way.  This
paucity of very large objects could be due to tidal destruction in the
cluster, but in any case we do not detect many galaxies abutting the
low-SB edge of our detection thresholds for $M_R<-12$.

\subsection{Galaxies and Globular Clusters with $M_R>-11.4$}

\label{globs}

As seen in Figure~\ref{lffig}, the Coma LF rises dramatically for
$M_R>-11.4$ with a slope of $\alpha=-2$. Other investigators are
reporting such upturns in cluster LFs, though at brighter magnitudes
than this (De~Propris \etal\ 1995).  How much of the upturn in object
counts faintward of $M_R=-11.4$ is attributable to globular clusters?
Is it possible that the Coma galaxy LF indeed turns sharply upwards at
these magnitudes?  If the number of globular clusters is negligible for
$M_R<-9.5$, then the galaxy counts for $-12.5<M_R<-9.5$ are indeed best
fit by a power law of index $\alpha=-2.0$.  If we assume to the
contrary that the $\alpha=-1.43$ power law fit to the $M_R<-11.4$ data
continues to describe the galaxy population at fainter levels, we would
predict that one third of our detections in the $-11.4<M_R<-10.4$
magnitude range were galaxies, with the remainder globular clusters.
As mentioned in \S\ref{resolved}, TV87 indeed resolve about one third
of the suspected Coma cluster members in this magnitude range, using a
small image with 0.6\arcsec~FWHM seeing.  This is consistent with the
$\alpha=-1.4$ extrapolation, but does not of course rule out the
existence of numerous extra dwarf galaxies too small even for Thompson
\& Valdes to resolve.

An independent estimate of the number of globular clusters expected in
our image can be made by assuming that the NGC~4874 globular cluster
system is a scale model of the M87 globular cluster system observed in
detail by McLaughlin, Harris, \& Hanes (1994, MHH94).  We first assume
that NGC~4874 is 5.5 times more distant than M87, which makes it 1.7
times more luminous in $V$ (de Vaucouleurs \etal\ 1991).  MHH94
tabulate the number of globular clusters by observed V magnitude;
assuming $V-R=0.5$ on average (Hopp, Wagner, \& Richtler 1995), we may
transform the M87 $V$ magnitudes to Coma-distance $R$ magnitudes by
adding 3.2~mag.   If the specific frequency of globular clusters is
similar to that of M87 (which is quite rich in globular clusters), then
there will be 1.7 times more total globular clusters around NGC~4874
than M87. We lastly need to correct for the fact that MHH94 count
globular clusters within a 6.8\arcmin\ radius of M87 (which would be
only 1.2\arcmin\ at the distance of Coma), while our field spans a more
extensive region around NGC~4874.  We will assume that the surface
density of globular clusters drops as $r^{-1.3}$ with distance from
NGC~4874, since this is the behavior we observe for the faintest
objects in our field.  It is furthermore consistent with the radial
gradients observed near M87 by MHH94, and with the $r^{-1.39\pm0.15}$
behavior observed for globular clusters within $\sim50$~kpc of the
center of the cD galaxy NGC~3311 (McLaughlin \etal\ 1995).  In this
case, there will be 1.9 times as many globular clusters in our field as
in the MHH94 field.  Given the magnitude shift and this multiplicative
factor we may scale the MHH94 data for M87 globulars to our field.  The
predicted globular cluster counts are shown as the dotted curve in
Figure~\ref{lffig}.

The simple scaling of the M87 globular cluster system to NGC~4874
predicts counts of globular clusters to be higher than the number of
Coma cluster objects we detect at $M_R>-12$.  This is particularly true
for $R\approx23$ ($M_R\approx-12$), where we can tell that most of our
objects are resolved, contrary to the prediction of the M87 globular
cluster scaling.  This is not a worry, because an acceptable range of
values for the Coma/Virgo distance ratio, the density gradient of the
globular clusters, or the specific frequency of clusters in NGC~4874
could easily accommodate a factor 2 change in the estimated globular
cluster density.  What this estimate does show, however, is that it is
unlikely that our faintest bins contain predominately dwarf galaxies.
Rather the globular clusters are probably a major component of our
faintest detections.

We refrain from attempting to constrain the parameters of the NGC~4874
globular cluster population---such as their luminosity function and
radial distribution---because our resolution does not permit a
satisfactory discrimination against dwarf galaxies.  From
Figure~\ref{lffig} it should be apparent that any effort to study dwarf
galaxies {\it or} globular clusters at $M_R\gtrsim-11$ must have
sufficient resolution to distinguish the two.  At the distance of the
Coma cluster, the refurbished HST is an excellent tool for studying the
globular clusters (see Baum \etal\ 1994), since they are unresolved and
easily detected, while the dwarf galaxies of $\gtrsim100$~pc size will
be quite well resolved.  Indeed because of their low surface brightness 
and the relatively
small and slow optics of the HST, it would take roughly 20~hours of
exposure time to detect the dwarfs with the HST.

\subsection{Mass Constraints on Dwarf Galaxies}

It is apparent from Figure~\ref{sizefig} that the total visible
luminosity from very faint galaxies ($-9.4>M_R>-15$) in the Coma core
is at most a few percent of the light emitted by the giant galaxies of
the region ($M_R\lesssim-18$).  We wish to estimate whether the {\it
mass} in the dE's is also negligible compared to the giants.  We can
make only the crudest estimates of the mass necessary to keep the dwarf
galaxies bound in the face of tidal forces from the cluster potential,
because we know little about the detailed shapes of the Coma dwarfs,
nor is the cluster potential well known.  Assume a dwarf galaxy of mass
$m$ and radius $r_t$ to be adrift at distance $r$ from the center of an
isothermal potential well with 1-dimensional velocity dispersion
$\sigma$.  Tidal forces will strip stars from the periphery of the
dwarf unless
\begin{equation}
\label{tidal}
 {Gm \over r_t^3} \ge {2 \sigma^2 \over r^2}.
\end{equation}
None of the quantities $r_t$, $\sigma$, or $r$ is particularly well
determined from our data, but we can make some estimates.  From
Figure~\ref{sizefig}, we find that the typical $R=23$ Coma dwarf
($M_R=-12$) fits an exponential scale length of $r_s=200$~pc.  Assume
that the galaxy extends for $k$ scale lengths before reaching the
truncation radius $r_t$.  If we assume that the Coma dwarfs have King
(1966) mass and light profiles with concentration parameters $2<c<6$,
then we find that $k\approx10$.  We have no direct evidence, however,
that the Coma dwarfs extend for 10 scale lengths.  We conservatively
assume that they are truncated at only 3 scale lengths.

The velocity dispersion of the Coma cluster is $\sim1000$~km~s$^{-1}$,
but the regions close to NGC~4874 might be well inside the core radius
of the cluster potential, where the tidal forces can be quite different
from the right-hand-side of Equation~(\ref{tidal}).  Instead we will
assume the dwarfs to be in an isothermal potential with
$\sigma=250$~km~s$^{-1}$, which is the measured central dispersion of
NGC~4874 (Faber \etal\ 1989).  The globular cluster system of M87 has
dynamics consistent with such a potential well within 50~kpc radius
(Merritt \& Tremblay 1993).  We observe the projected dwarf galaxy
density to be rising at distances $\lesssim1\arcmin$ from NGC~4874, so
we assume $r\approx30h_{75}^{-1}$~kpc.  Equation~(\ref{tidal}) can be
rearranged as
\begin{equation}
\label{ml1}
\left( {m \over L} \right) \ge 2.3h_{75}\, 
	\left( {m_\odot \over L_\odot} \right)\,
	10^{0.4(R-23)}
	\left( k \over 3 \right)^3
	\left( r_s \over 200\,{\rm pc} \right)^3
	\left( \sigma \over 250\,{\rm km\,s}^{-1} \right)^2
	\left( r \over 30\,{\rm kpc} \right)^{-2},
\end{equation}
where the luminosities are in $R$ band.  The Local Group dwarf
spheroidals have $m/L$ in the range 10--100 (FB94 and references
therein), and would thus probably survive without serious tidal
stripping if dropped into the NGC~4874 potential well.  There is no
need for the Coma dwarfs to have extraordinary $m/L$ values to be
long-lived, and thus they likely contribute negligibly to the total
mass of the cluster (assuming that the radial density gradient of the
dwarfs is at least as steep as that of the giants throughout the
cluster).  This conclusion is, however, strongly dependent on the
nature of the gravitational potential and tidal field in the inner
100~kpc of the cluster.  The mass distribution in this region is quite
uncertain, so these conclusions could easily require revision.

Note that the $m/L$ value required to maintain integrity in the tidal
field scales as $r_s^3/L$.  At a given magnitude, tidal destruction
could produce an upper limit to the $r_s$ (or equivalently a lower
$\mu_0$ limit) distribution of dwarfs in Coma.

Dwarf galaxies in the Coma core are heated by encounters with giant
galaxies due to tidal forces.  The dwarf galaxies could become unbound
if enough energy is injected over the $\sim5$~Gyr since the creation of
the Coma cluster.  We calculate that the $m/L$ ratios necessary to
stave off this effect are an order of magnitude lower than the estimate
in Equation~(\ref{ml1}), and can be safely ignored.

\section{Implications and Conclusions}

We have measured the LF in the Coma cluster core to fainter absolute
magnitudes ($M_R=-11.4$, $L=2h_{75}^{-2}\times 10^6 L_\odot$) than any
other LF study of which we are aware, and with a broader sensitivity in
surface brightness (the recent De~Propris \etal\ [1995] study is to
similar depth as ours).  We have found that an $\alpha=-1.4$ power law
describes the LF down to luminosities typical of the Local Group dwarf
spheroidal galaxies.  These most extreme Coma cluster galaxies have
sizes comparable to Local Group galaxies at a given absolute
magnitude.  This should be a useful constraint for theories of galaxy
formation.  We have, however, performed this measurement in a most
extreme environment, one of the densest regions of the nearby
Universe---there are $10^5$--$10^6$ galaxies per Mpc$^3$ in the Coma
core, compared to 10-100 per Mpc$^3$ in the Local Group.  This
complicates comparison to the general field LF because of the many
additional processes which could have influenced the development of the
Coma dwarfs, such as the increased pressure and tides in the cluster
environment.  The significance of these results to theories of galaxy
or cluster formation depends upon the evolutionary history of the Coma
dwarfs:  are their formation and evolution identical to those of field
dwarfs, merely being concentrated for our convenience in the core of
the cluster?  Or, at the other extreme, are they formed by processes
that exist solely in the cluster core, giving them little relation to
faint Local Group or field galaxies?  We first discuss the possibility
that Coma and field dwarfs are very similar, and then move on to
scenarios for the Coma dwarfs that are increasingly disparate from
those of field galaxies.  We refer the reader also to the FB94 review
for discussions of evolutionary scenarios for dE galaxies; the
formation of dwarf galaxies (indeed all galaxies) is poorly
understood---it is not even clear what are the dominant physical
processes---so we will not go into depth on any particular hypothesis.

\subsection{Coma Dwarf Population Similar to the Field Population?}

A pleasingly simple interpretation would be that the dwarf galaxies in
the Coma cluster core are an entombed, concentrated sample of field
dwarf galaxies.  The giant galaxies in the Coma cluster are extremely
atypical compared to the field, being nearly completely devoid of late
types.  Dwarf galaxies, however, are much less massive and thus less
subject to dynamical friction than giants.  Furthermore they are
smaller, and could (depending on the scaling of $m/r_t^3$ with $L$) be
less subject to tidal disruption than the giants, with our simplest
calculations indicating that Local Group dwarfs would not be severely
harmed by the tidal field around NGC~4874.  Thus a dwarf galaxy
probably has a better chance than a giant of inhabiting the cluster
environment without substantial merging or destruction.  It is likely,
however, that dwarf galaxies in the Coma cluster would be stripped of
gas (FB94), and thus the episodic star formation history that seems
characteristic of the Local Group dwarfs (see FB94 for review) would
have been truncated for the Coma dwarfs soon after the development of
the hot intra-cluster medium.  This of course has been advanced as an
explanation of the apparent deficit of dwarf irregular galaxies in
clusters relative to the field, as discussed by FB94.  They also note,
however, the dominance of gas-stripped dE's over gas-rich irregular
dwarfs in regions of Virgo where stripping should {\it not} be
important, indicting some internal mechanism as the agent of gas
stripping, rather than ram pressure.  Thus it is possible that the Coma
cluster medium has had little evolutionary effect on the dwarfs we
observe to be resident there.

In this most simplistic case, the Coma dwarfs are a typical population
of low-luminosity galaxies, save that they have had no recent star
formation episodes to significantly perturb their luminosities.  The
agreement between our faint-end LF slope and those from the Virgo
studies (SBT85 and the LSB extension of Impey, Bothun, \& Malin 1988)
supports the idea of a universal mass function (or more precisely,
universal stellar mass function) for dwarf galaxies.  Of course should
the much steeper cluster LF slopes measured by De~Propris \etal\ (1995)
be confirmed, we would have to abandon this point of view.

The main problem with the idea of a universal faint-end LF is that
field surveys yield shallower slopes:  $\alpha=-0.97\pm0.25$ from
Loveday \etal\ (1992); $\alpha=-1.0\pm0.2$ from Marzke, Huchra, \&
Geller (1994); $\alpha=-1.1$ from Ellis \etal\ (1995).  In the Local
Group the LF may be determined for even fainter galaxies than in our
study:  van~den~Bergh (1992) fits $\alpha=-1.1$ to the Local Group
galaxy LF for $M_V<-7.6$.  With new Local Group dwarfs being discovered
on an annual basis, however, it is possible that this slope will
steepen with time.  Babul \& Rees (1992) proposed that dwarf galaxies
are born with common mass functions in both cluster and field
environments.  Field dwarfs might effectively blow themselves up in
star formation incidents, but cluster dwarfs would remain confined by
the pressure of the intra-cluster medium.  In this view the field LF
vs.\ cluster LF dichotomy is real and due to the demise of field
dwarfs.  We have found a faint-end LF in Coma similar to that in Virgo,
and Thompson \& Gregory (1993) also report a dwarf-to-giant ratio
similar in Coma and Virgo, despite the former being a denser
environment.  Thus if the Babul \& Rees scenario is correct, then the
confinement effect may ``saturate'' at Virgo densities---the additional
pressure in Coma preserves no more dwarfs, and $\alpha\approx-1.4$
represents the ``intrinsic'' dE LF.  Countering this view somewhat is
the observation that Coma cluster dwarf galaxies have similar
sizes---and perhaps similar masses---to the Local Group dwarfs at
comparable magnitudes.  If dE evolution is strongly controlled by local
pressure, would we expect Local Group and Coma cluster objects to be
this similar?

Another school of thought on the field/cluster LF dichotomy is that the
differences are primarily due to selection effects.  In particular, it
is suggested that cluster LF studies are generally done with deeper
images than are nearby field surveys, resulting in omission of LSB
galaxies from the field surveys.  The field survey of Ellis
\etal\ (1995), for example, selects target galaxies from a survey with
limiting surface brightness 26.5~$b_J$~mag~arcsec$^{-2}$, substantially
shallower than the Virgo, Fornax, and Coma cluster surveys discussed
here.  Deep field redshift surveys seem to show an increase in the LF
slope at higher redshifts (Ellis \etal\ 1995; Eales 1993), which could
likewise be due to the fact that deep redshift survey targets are
selected from deeper photographic or CCD exposures with more
sensitivity to LSB objects.  Some steepening in the LF even at fixed SB
sensitivity does, however, seem to be indicated by the Ellis
\etal\ (1995) survey, since their $z\approx0$ and their $z\approx0.3$
LFs are constructed from the same parent survey, with a single SB
threshold, yet they find a steeper LF at higher $z$.  Ferguson \&
McGaugh (1995) also suggest that local field LF slopes are depressed by
failure to count LSB galaxies.  This may be evidenced by the excess
(over the $\alpha=-1.0$ prediction) of nearby $-16<M_{\rm Zw}<-13$
galaxies seen in the CfA survey by Marzke, Huchra, \& Geller (1994).
Marzke \etal\ (1994) attribute this excess to a population of Sm-Im
galaxies with $\alpha=-1.87$; the relation of such galaxies to cluster
dE's is unclear.  Sprayberry \etal\ (1995) report observations of the
field LF that have the faint-end slope significantly increased when
they make extra effort to reduce surface-brightness biases.

Gronwall \& Koo (1995) derive a $z=0$ field LF for galaxies by
requiring a best fit to faint galaxy count, color, and redshift survey
data.  They claim that most of these data can be explained by allowing
the local LF to steepen to $\alpha\approx-1.5$ at $M_{B_J}>-17$, where
local information is sparse.  Such behavior would be consistent also
with our Coma cluster LF.  Their derived LF is consistent with the
Loveday \etal\ (1992) LF, but may disagree with the deeper Ellis
\etal\ (1995) $M_{B_J}<-15$ LF.

Should the field LFs somehow be reconciled with the Coma or Virgo LFs,
or the Gronwall \& Koo LF, we would be led to suspect some internal
process ({\it e.g.\/} supernova heating) as the driver of dwarf galaxy
evolution rather than environmental variables such as pressure or
ionizing radiation field.

\subsection{Inhibition or Destruction of Cluster Dwarfs}

While the Babul \& Rees (1992) scenario invokes the intra-cluster
medium to {\it increase} the number of dwarfs in clusters over the
field counts, there are of course processes that could {\it reduce} the
number of cluster dwarfs, especially within 100~kpc of the Coma cluster
core.  While our simple calculations suggested that tidal forces are
not important, a more interesting test will be to compare the size
distributions of the Coma dwarfs to those in the Virgo cluster.  In a
further publication, after re-analyzing the most diffuse objects in our
image, we will compare the maximum sizes of galaxies in the Coma core
to those found in Virgo.  Should the latter be larger at a given
magnitude, it would suggest tidal forces are indeed at work in Coma.
For now we simply note that the $M_R\approx-12$ galaxies in the Coma
cluster do not appear to be depleted near the core, as might be
expected were tidal forces destroying most dwarfs.

\subsection{Satellites of NGC 4874}

The dwarf galaxies in our image show a strong concentration toward the
giant elliptical NGC~4874, which may not be the dynamical center of the
Coma cluster.  It may be incorrect to think of these dwarfs as
belonging to the cluster; a more appropriate description may be as
satellites of NGC~4874.  The spatial distribution of dwarfs around
NGC~4874 is consistent with that observed for satellites of field
ellipticals (Vader \& Sandage 1991) and of field spirals (Zaritsky
\etal\ 1993), albeit far richer.  The neighborhoods of giant
ellipticals may be particularly fertile for production of dwarf
galaxies.  It would be interesting to see whether NGC~4889, or
NGC~4881, which are giant Coma cluster ellipticals with apparently
lower specific frequencies of globular clusters than NGC~4874 (Harris
1987; Baum \etal\ 1995), also have fewer dwarf galaxies in their
vicinities.  Note in Figure~\ref{radfig}, however, that the brightest
galaxies in Coma may have an equally strong concentration toward
NGC~4874, though the number of objects in the nearest bin is small
(5).  Note further that we did not detect any concentration of dwarf
galaxies around the other elliptical galaxies in our Coma field---but
these are each at least ten times less luminous than NGC~4874.  If the
dwarf galaxies which we have detected in the Coma core are members of
an enhanced satellite population, then the comparison to the field LF
is further complicated, and environmental mechanisms are of course
implicated in the formation and evolution of these galaxies.

In this context it is worth noting that the diffuse light, globular
cluster density, and dwarf galaxy density seem to have similar radial
structure on the outskirts of NGC~4874.  McLaughlin \etal\ (1995) note
that all the current data on globular cluster populations around giant
ellipticals are consistent with the idea that globular clusters and cD
envelopes (as opposed to the ``bodies'' of cD galaxies, which follow an
$r^{-1/4}$ profile) always have common structure.  NGC~4874 is
classified as a cD galaxy by Schombert (1988), and our data are taken
at large enough distances from its center that the envelope light is
dominant, by his definitions.  If the globular clusters are part of the
same system as the diffuse envelope, then our data may be suggesting
that the formation of the dwarf galaxies is likewise integrally
connected with the mysterious origin of the cD envelope.

\subsection{Dwarfs as Shards}

In the above subsections we assumed that the objects we detect in a
cloud around NGC~4874 are long-lived and predate the cluster.  Neither
need be true---the dwarfs could be pieces of larger galaxies destroyed
by the cluster tidal fields or through interaction with NGC~4874.  They
could also be in the process of dissolution.  Note that the dwarf
galaxies show roughly the same spatial distribution as the diffuse flux
around NGC~4874, but with only a few percent as much total luminosity.
If dwarf ``galaxies'' are constantly being formed and dissolved near
NGC~4874, then each must typically live at least a few percent of the
age of the system, else there will be too much diffuse light left
over.  We would perhaps expect, however, to see large numbers of very
extended dwarfs were there a continual process of dissolution, and we
do not.    If the stellar contents of the dwarfs were to fade before
becoming part of the diffuse light, then the dwarfs could be even
shorter-lived, and we might not detect them in their most diffuse
state.  Color or spectral information would fairly quickly tell us
whether in fact these objects are transient ($\lesssim10^7$ year)
starburst phenomena.

The similarity between the diffuse light and the dwarf galaxy gradient
suggests a common, unknown origin.  It is often suggested that cD
galaxy halos are the remains of galactic cannibalism.  If the dwarf
galaxies are shards of NGC~4874's victims, then their colors should
resemble those of the old populations of giant galaxies.  The colors of
field and cluster dE's more closely resemble those of metal-poor
globular clusters.

A further point worth noting about the diffuse light is that it is, in
a nutshell, diffuse.   We calculate that the visible ($R$) component of the
hot intracluster bremsstrahlung radiation [detected in X-rays, {\it cf.\/}
White, Briel, \& Henry (1993)]
is several orders of magnitude
weaker than the detected $R$-band flux, so that most of the $R$-band should be starlight.  Our sensitivity is such that any unresolved
clumps brighter than $M_R=-9.4$, or $3\times10^5\,L_\odot$, would be
detected.  Yet the total flux in such lumps is still only a few percent
of the diffuse flux.  Even if we were to extrapolate the $\alpha=-1.4$
LF to zero-luminosity galaxies, we would not have enough flux to make
up the diffuse light.  If the diffuse light is lumpy, the lumps must be
quite small.  Scheick \& Kuhn (1994) similarly conclude, from a search
for surface-brightness fluctuations in the diffuse light of the cluster
Abell~2670, that the typical unit of luminosity in the diffuse light
must be $\le3\times10^3\,L_\odot$.  Thus if the diffuse light is formed
from disrupted galaxies, the remnants are very effectively dispersed.
It would seem odd for $10^6\,L_\odot$ objects to be left behind, with
scale sizes of $\sim200$~pc just like Local Group dSph's. The origin of
the cloud of diffuse light, globular clusters, and dwarf galaxies
surrounding NGC~4874 begs further investigation, especially since the
flux of starlight in the diffuse component is comparable to the total
amount in galaxies of any size within the central 100~kpc.

\subsection{Implications for the Coma Cluster}

Finally, the observations presented in this paper have implications for
the Coma cluster as a whole. As stated earlier, the implied total
amount of mass in previously undetected galaxies and globular clusters
is insignificant compared to that already postulated from the brighter
galaxies in the cluster (assuming mass-to-light ratios consistent with
those measured in the Local Group dwarfs). The Coma dwarfs would need
to have mass-to-light ratios several orders of magnitude greater than
any previously measured dwarf galaxy to have a significant impact on
the total mass of the cluster.  We have surveyed and catalogued
all objects in the center of Coma having luminosities of globular clusters
or brighter, plus we have measured the
diffuse halo light. It is thus highly unlikely that there remains
undetected, visibly luminous matter
sufficient to close the cluster.  White \etal\ (1993) have
carried out a full inventory of the x--ray gas mass in Coma and also
find this to be grossly insufficient to close the cluster.  Unless
there is a large population of extremely faint massive objects in Coma,
we are left once again with the hypothesis that Coma and clusters in
general are dominated by non-luminous mass.  Recent observations of
microlensing by massive compact objects in our own galaxy suggest that
only $\sim20$\% of the total galaxy mass can be accounted for in
extremely faint baryonic objects (Gates \etal\ 1995).  Thus, even if
every one of the Coma galaxies had a population of MACHOS, it would not
close the cluster.  These circumstances continue
to suggest the presence of non-baryonic dark matter in the cluster.

The discussion above also has bearing on recent claims of a baryon
crisis in Coma. White \etal\ (1993) have suggested that the amount of
baryons seen in Coma is in conflict with the bounds set from
nucleosynthesis. Our observations have shown that the amount of baryons
in faint, previously unknown objects is inconsequential to this
argument. The extra amount of baryonic mass added to their inventory of
the cluster is well within their quoted error estimates.

\section{Summary and Future Directions}

Our census of the population of galaxies in the Coma cluster has
produced a few basic facts about extremely faint galaxies:  first, the
LF in the extraordinarily dense Coma cluster core is similar in shape
to that measured in the Virgo cluster by SBT85, though our survey
extends to fainter magnitudes, and is sensitive to both higher and
lower surface brightness dwarf galaxies.  A power law with
$\alpha=-1.4$ describes both populations fairly well, down to
$M_R=-11.4$ in our study.  Further extrapolation of this law to
$M_R=-9.4$ is consistent with our data and those of TV87, in the regime
where globular clusters are easily confused with dwarf galaxies.  This
LF is in conflict with field LF studies and the LF of known Local Group
members, both of which seem to indicate shallower faint-end slopes.  On
the other hand, the sizes and masses of the faintest detected Coma
galaxies seem roughly consistent with those of Local Group dwarf
galaxies of similar luminosity, and there is as yet little evidence
that the Coma dwarfs have had their evolution severely affected by
their proximity to NGC~4874.  The $M_R\sim-12$ dwarf galaxies follow
the diffuse light and globular cluster populations in a strong
concentration near NGC~4874, with all three increasing in density as
$r^{-1.3}$ towards this cD galaxy.  This gradient is similar to that
observed for globular clusters and dwarf galaxies around giant
ellipticals in the field, so it may be that we are viewing a population
that was born with NGC~4874.

The collection of these data is not a severe challenge for modern
instrumentation, and extension of this work would be straightforward.
Indeed further such observations of Coma and other distant clusters are
in progress ({\it e.g.\/} Secker \& Harris 1994; De Propris
\etal\ 1995).  Our caution to future investigators is that the quality
of background correction is both improved and more easily assessed if
control fields covering several times the area of the target fields are
obtained.  Many newer telescopes routinely obtain seeing of
0.8\arcsec\ or better; in such conditions, the distinction between
galaxies and globular clusters would be easier, and the Coma galaxy LF
could be extended another magnitude or so fainter.  HST images could be
used to identify the globular clusters and remove them from the galaxy
LF (HST images of control fields would be necessary as well).

Are the Coma cluster dwarfs similar to Local Group or field dE's?  A
measure of the colors of the Coma dwarfs is clearly called for. Do they
match those of field, Local Group, or Virgo dwarfs?  Is there any sign
of recent star formation in these objects?  This seems unlikely, but
color information could rule out any sort of transient nature for these
structures.  Colors as blue as the Virgo dE's (Caldwell \& Bothun 1987)
would rule out an origin as remnants of destroyed larger galaxies.  In
measuring these colors, extensive color information on control fields
would of course have to be obtained, since the background subtraction
techniques we have employed would have to be implemented on the
color-magnitude plane.

Spectroscopic information will be difficult to obtain given the faint
apparent magnitudes, but a measure of the velocities of these many
dwarfs could enlighten us as to the dynamical state of the inner
regions of Coma, where there are not enough giants to learn about the
dynamics.  Of course spectral information would also tell us about the
stellar populations and metal content of these dwarfs.

The Coma cluster is useful in that it contains two cD galaxies and a
third giant elliptical (Schombert 1988).  Replicating our study near
NGC~4889 and NGC~4881 would enlighten us as to whether the dwarfs are
truly associated with the bottom of the cluster potential well, or
whether they are found in large numbers in the vicinity of all giant
ellipticals, or only near cD galaxies.  Invaluable clues to their
origin would result.  It may be difficult to search for the
$M_R\sim-12$ galaxies on the outskirts of Coma, as their numbers may
become too sparse to detect in the midst of background galaxy density
fluctuations.

We thus hope that our Coma census will be a cornerstone for these
future studies, which will build up the edifice of dwarf galaxy
formation theory.  Thousands of these galaxies are accessible to
observations by modern instrumentation, which will help determine the
extent to which cluster dE's and field dwarf galaxies share a common
structure or history.

\acknowledgements
Thanks to P. Teague for assistance in the Coma observing run, and to P.
Guhathakurta for allowing us to commandeer images from other projects
for use as control fields.  We thank N. Metcalfe for providing an
electronic version of the GMP83 galaxy catalog.  GMB was generously
supported by AT\&T Bell Laboratories and the Bok Fellowship from
Steward Observatory during the tortuously extended duration of this
work.  GMB also thanks E. Olszewski, N. Caldwell, C. Impey, and M.
Mateo for discussions of dwarf galaxies.  MPU thanks A.~Sandage for
discussions at the inception of this project, and Northwestern
University and NASA for partial support.

\clearpage
\section*{Figure Captions}

\begin{figure}[h]
\caption[dummy]{
Deep CCD Image of the Coma Core.  This image is approximately
7.5\arcmin\ square; North is up, East is to the left, and the intensity
has been logarithmically scaled to increase the dynamic range.  NGC~4874
is just off the N edge at the right half of the image; the diffuse light
gradient from NGC~4874 is evident across the entire field.
}
\label{rawpic}
\end{figure}

\begin{figure}[h]
\caption[dummy]{
Cleaned Image of the Coma Core.  The same CCD field as shown in
Figure~\protect{\ref{rawpic}} is shown here after subtraction of the
diffuse light signal and the elliptical models of the brightest
objects.  Areas with large residuals to these models are masked off.
This image is linearly scaled, and thousands of faint objects are now
visible, with a marked gradient toward NGC~4874.}
\label{cleanpic}
\end{figure}

\begin{figure}[h]
\caption[dummy]{
Magnitude-Resolution Distribution of Coma Field.  Resolution
parameter \res\ and total magnitude in the $R$ band are shown for all objects
in the Coma field catalog.  Objects in the region with $R<21$ and $\res<0.8$
are assumed to be foreground stars since they are unresolved and distinct
from the galaxy population.
}
\label{resfig}
\end{figure}

\begin{figure}[h]
\caption[dummy]{
Object Counts and Completeness.  The upper plot shows the
detection efficiency for unresolved objects as a function of magnitude,
as derived from the Monte Carlo tests described in \protect{\S\ref{monte}}  
The heavy
solid line is for the Coma field, while the five dotted lines are for
the five control fields.  The lower panel shows the number of objects
detected per magnitude for the Coma field as the histogram; stars
are excluded from the counts for $R<21$.  The
triangles with error bars give the estimated background counts $\tilde N_{bg}$
and the uncertainties $\sigma_{bg}$ derived from the control fields.
The dashed line shows the $dN/dm\propto {\rm dex}(-0.39R)$ power-law behavior
expected of the background counts (Tyson 1988), and the dotted line
shows the star counts expected in the model of Ratnatunga (1993).}
\label{compfig}
\end{figure}

\begin{figure}[h]
\caption[dummy]{
Luminosity Function of the Coma Cluster Core.  The filled circles show
the number $N_{cl}$ of objects present in the Coma cluster core within
the area of our CCD image.  The open triangles are our best estimate of
the number of resolved objects in these bins; in brighter bins,
essentially all objects are resolved, and unresolved in fainter bins.
The open circles show the number of objects per magnitude within
8\arcmin\ of NGC~4874 from the GMP83 photographic catalog.  All points
have also been corrected for background counts and incompleteness, and
scaled to the 52.2~arcmin$^2$ area of the CCD field.  The straight line
is the best fit power-law to the luminosity function over the range
$15.5<R<23.5$.  This best fit has $dN/dL\propto L^\alpha$ with
$\alpha=-1.42\pm0.05$.  If the measured globular cluster population
around M87 were moved to NGC~4874 as described in
\protect{\S\ref{globs}}, we would have detected the number of globular
clusters shown by the dotted line.  Thus it is likely that the majority
of the $R>23.5$ objects in the cluster are globular clusters rather
than galaxies.  The absolute magnitude scale on top assumes a distance
modulus of 34.9 to the Coma cluster.
}
\label{lffig}
\end{figure}

\begin{figure}[h]
\caption[dummy]{
Resolution Distribution of Coma Cluster Members.  The histogram in each
panel shows, for the indicated magnitude range, the number of objects
in the Coma field as a function of measured resolution parameter.  The
background contribution in each bin has been estimated from the control
fields and subtracted; the error bars show the uncertainty due to
background count fluctuations.  The dotted curves show the
distributions of resolution parameter expected from a population of
unresolved objects.  For the brightest two bins shown here there must
be a sizable contribution from resolved objects, whereas the fainter 4
bins are consistent with the unresolved distribution.
}
\label{slicefig}
\end{figure}

\begin{figure}[h]
\caption[dummy]{
Sizes of Faint Coma Members.  This plot shows the distribution of
magnitude and exponential scale length for the very faint objects we
detect in the Coma cluster core.  The heavily shaded region shows the
typical sizes of Coma cluster galaxies with $22.5<R<23.25$.  There are
relatively few Coma galaxies in this magnitude range with scale lengths
above or below the shaded range.  For $R>23.25$, our ability to
determine the sizes of the population is diminished, but we detect many
objects.  The lightly hatched region outlines the ranges of galaxy
magnitude and size which these objects could have.  The heavy line at
right is the locus of 50\% detection efficiency; we are seriously
incomplete to the right of this line.  The Local Group dwarf spheroidal
galaxies (save Fornax, which is too bright) are labelled at their
respective locations in this magnitude, size plane.  The sizes of the
extremely faint Coma cluster galaxies seem to be similar to the sizes
of Local Group galaxies at comparable luminosity.  Our sensitivity is
sufficient to detect most of the Local Group dSph's at the distance of
the Coma cluster.
}
\label{sizefig}
\end{figure}

\begin{figure}[h]
\caption[dummy]{
Radial Distributions of Coma Cluster Members.  The mean $R$ surface
brightness contributed by Coma cluster members of various magnitudes is
shown as a function of distance $r$ from NGC~4874.  The legend lists
the magnitude ranges corresponding to the various symbols; in all
cases, the error bars reflect uncertainties in the background levels,
as described in the text.  The dotted lines trace a surface brightness
dropping as $r^{-1.3}$.  The diffuse light and all of the discrete
objects are consistent with this radial dependence, except that
galaxies of moderate brightness ($-17.4<M_b<-14.9$, filled triangles)
seem to flatten for $r\lesssim 5\arcmin$.  Somewhat fainter galaxies
($-14.4<M_R<-12.4$, squares) may also have a central deficit, but the
bright galaxies, extremely faint galaxies ($M_R>-12.4$), and globular
clusters are consistent with the power law across the sampled range in
$r$.  The total light contributed by very faint galaxies is small
compared to the surface brightness in diffuse light or in galaxies at
the brightest bin.  The brightness of the night sky in $R$ band is
20.3~mag~arcsec$^{-2}$.
}
\label{radfig}
\end{figure}

\clearpage
\begin{planotable}{ccccccccc}
\tablewidth{0pt}

\tablecaption{Fields Observed}
\tablehead{
\colhead{Name} & 
\colhead{RA} & 
\colhead{Dec.} &  
\colhead{$l$} &  
\colhead{$b$}  &
\colhead{Area} &
\colhead{FWHM\tablenotemark{a}} &
\colhead{Noise Density\tablenotemark{a}} &
\colhead{$A_R$} \nl
 & 
\multicolumn{2}{c}{(1950)} & & & 
 (arcmin$^2$) & (arcsec) & ($R$ mag arcsec$^{-1}$)
 & (mag) 
}
\tablenotetext{a}{After degradation to match Coma.}
\startdata
Coma & $12^{\rm h}57^{\rm m}17^{\rm s}$ & $+28\deg09\arcmin35\arcsec$ &
  56\deg & $+88\deg$ & 52.2 & 1.31 & 28.07 & 0.00 \nl
SA68 & $00^{\rm h}14^{\rm m}53^{\rm s}$ & $+15\deg29\arcmin27\arcsec$ &
  111\deg & $-46\deg$ & 54.2 & 1.31 & 27.94 & 0.02 \nl
0427 & $04^{\rm h}27^{\rm m}56^{\rm s}$ & $-36\deg24\arcmin09\arcsec$ &
  238\deg & $-43\deg$ & 52.9 & 1.30 & 28.02 & 0.00 \nl
0909 & $09^{\rm h}09^{\rm m}57^{\rm s}$ & $-07\deg38\arcmin18\arcsec$ &
  238\deg & $+26\deg$ & 53.9 & 1.28 & 28.09 & 0.06 \nl
Ser1 & $15^{\rm h}13^{\rm m}13^{\rm s}$ & $+00\deg49\arcmin46\arcsec$ &
  0\deg & $+45\deg$ & 54.1 & 1.33 & 27.98 & 0.12 \nl
Her1 & $17^{\rm h}20^{\rm m}39^{\rm s}$ & $+50\deg06\arcmin19\arcsec$ &
  76\deg & $+35\deg$ & 54.7 & 1.32 & 28.12 & 0.01 
\end{planotable}

\begin{planotable}{crrrrrrr@{$\pm$}rr@{$\pm$}l}
\tablewidth{0pt}
\tablecaption{Coma CCD Object Counts}
\tablehead{
\colhead{$R$} &
\colhead{$N_C$} &
\colhead{$\tilde N_{bg}$} &
\colhead{$N_{det}$} &
\colhead{$\sigma_{det}$} &
\colhead{$N_{cl}$} &
\colhead{$\sigma_{cl}$} &
\multicolumn{2}{c}{$dN_{cl}/dm$} &
\multicolumn{2}{c}{$dN_{res}/dm$} 
}
\startdata
15.50--16.50 &    7  &    0.00 &     7.00 &     0.00 &    7.0 &     2.6 & 
   7.0 &  2.6  & \multicolumn{2}{c}{$\cdots$} \nl 
16.50--17.50 &   11  &    1.75 &     9.25 &     0.90 &    9.2 &     3.2 &  
  9.2 &  3.2  & \multicolumn{2}{c}{$\cdots$} \nl
17.50--18.50 &   11  &    3.27 &     7.73 &     3.01 &    7.7 &     4.1 &  
  7.7 &  4.1  & \multicolumn{2}{c}{$\cdots$} \nl
18.50--19.50 &   17  &    9.96 &     7.04 &     4.35 &    7.2 &     5.2 &  
  7.2 &  5.2  & \multicolumn{2}{c}{$\cdots$} \nl
19.50--20.50 &   43  &   27.18 &    15.82 &     9.76 &   16.2 &    10.8 & 
  16.2 & 10.8  & \multicolumn{2}{c}{$\cdots$} \nl
20.50--21.50 &   87  &   74.29 &    12.71 &    15.50 &   13.1 &    16.4 &  
 13.1 & 16.4 &  27 & 8\nl
21.50--22.50 &  171  &  168.34 &     2.66 &    35.81 &    2.1 &    37.0 &  
  2.8 & 37.0 &  23 & 15\nl
22.50--23.00 &  170  &  142.22 &    27.78 &    25.43 &   29.1 &    27.2 &  
 58.1 & 54.3 &  69 & 16\nl
23.00--23.50 &  272  &  206.18 &    65.82 &    22.19 &   71.5 &    25.7 &  
 143.1 & 51.4 & 100 & 28\nl
23.50--24.00 &  395  &  311.33 &    83.67 &    26.61 &   92.2 &    31.0 &  
 184.3 & 62.0 &  88 & 88\nl
24.00--24.50 &  532  &  381.58 &   150.42 &    27.52 &  179.5 &    35.9 &  
 359.0 & 71.9 & \multicolumn{2}{c}{$\cdots$} \nl
24.50--25.00 &  658  &  476.78 &   181.22 &    32.34 &  242.3 &    46.8 &  
 484.6 & 93.7 & \multicolumn{2}{c}{$\cdots$} \nl
25.00--25.50 &  739  &  522.40 &   216.60 &    71.72 &  340.6 &   115.1 & 
 681. & 230. & \multicolumn{2}{c}{$\cdots$} \nl
25.50--26.00 &  574  &  446.63 &   127.37 &    57.63 &  471.7 &   217.5 &  
 943. & 435. & \multicolumn{2}{c}{$\cdots$}
\end{planotable}

\begin{planotable}{cr}
\tablewidth{200pt}
\tablecaption{Coma Photographic Object Counts}
\tablehead{
\colhead{$R$} &
\colhead{$dN_{cl}/dm$\tablenotemark{a}}
}
\tablenotetext{a}{Object counts within 8\arcmin\ of NGC~4874 have been
corrected for background and scaled to the area of the CCD field.}
\startdata
10--11 & $0.24\pm0.24$ \nl
11--12 & $0.24\pm0.24$ \nl
12--13 & $-0.03\pm0.24$ \nl
13--14 & $1.85\pm0.68$ \nl
14--15 & $3.89\pm0.99$ \nl
15--16 & $3.60\pm0.96$ \nl
16--17 & $4.83\pm1.15$ \nl
17--18 & $8.05\pm1.57$ 
\end{planotable}


\begin{references}
Babul, A. \& Rees, M. 1992, \mnras, 255, 346

Baum, W. A. \etal\ 1994, \baas, 26, 1398

Biviano, A. \etal\ 1995, \aap\ (submitted)

Bothun, G. D., Impey, C. D., \& Malin, D. F. 1991, \apj, 376, 404.

Burstein \& Heiles 1982, \aj, 87, 1165

Caldwell, N., Armandroff, T. E., Seitzer, P., \& Da Costa, G. S. 1992, 
\aj, 103, 840

Caldwell, N., \& Bothun, G. D. 1987, \aj, 94, 1126

De Propris, R., Pritchet, C. J., Harris, W. E., \& McClure, R. D. 1995,
\apjl\ (in press)

de Vaucouleurs, G., \etal\ 1991, Third Reference Catalogue of Bright
 Galaxies, Springer-Verlag, New York

Disney, M. 1973, \apj, 181, L55

Driver, S. P., Phillipps, S., Davies, J. I., Morgan, I., \& Disney, M. J. 1994,
 \mnras, 268, 393

Dow, K. L., \& White, S. D. M. 1995, \apj, 439, 113

Eales, S. J. 1993, \apj, 404, 51.

Ellis, R.S., Colless, M.,
 Broadhurst, T.J., Heyl, J.S. \& Glazebrook, K. 1995 (in preparation)

Faber, S. M. \etal\ 1989, \apjs, 69, 763.

Ferguson, H. C. 1993, \mnras, 263, 343

Feguson, H. C. \& Binggeli, B. 1994, A\&ARev, 6, 67 (FB94)

Ferguson, H. C., \& McGaugh, S. S. 1995, \apj, 440, 470

Ferguson, H. C., \& Sandage, A. 1988, \aj, 96, 1520 (FS88)

Ferguson, H. C., \& Sandage, A. 1991, \aj, 101, 765

Gates, E. I., Gyuk, G., \& Turner, M. S. 1995, \prl, (submitted)

Godwin, Metcalfe, \& Peach 1983, \mnras, 202, 113 (GMP83)

Gronwall, C., \& Koo, D. C. 1995, \apjl, 440, L1

Harris, W. E. 1987, \apjl, 315, L29

Hopp, U., Wagner, S. J., \& Richtler, T. 1995, \aap\ (in press)

Impey, C., Bothun, G., \& Malin, D. 1988, \apj, 330, 634

King, I. R. 1966, \aj, 71, 64

King, I. R. 1971, \pasp, 83, 199

Loveday, J., Peterson, B. A., Efstathiou, G., \& Maddox, S. J. 1992,
 \apj, 390, 338

Lugger, P. M. 1986, \apj, 303, 535

Marzke, R. O., Huchra, J. P., \& Geller, M. J. 1994, \apj, 428, 43.

Mazure, A., Proust, D., Mathez, G., \& Mellier, Y. 1988, \aaps, 76, 339

McLaughlin, D. E., Harris, W. E., \& Hanes, D. A. 1994, \apj, 422, 486 (MHH94)

McLaughlin, D. E., Secker, J., Harris, W. E., \& Geisler, D. 1995, \aj, 109, 
  1033

Merritt, D., \& Tremblay, B. 1993, \aj, 106, 2229

Ratnatunga, K. 1993 (private communication).

Reed, B. C., Hesser, H. E., \& Shawl, S. J. 1988, \pasp, 100, 545

Romani, R. W. \& Maoz, D. (1992), \apj, 386, 36

Sandage, A., Binggeli, B., \& Tammann, G. A. 1985, \aj, 90, 1759. (SBT85)

Schombert, J. M. 1988, \apj, 328, 475

Secker, J. \& Harris, W. E. 1994, \baas, 26, 1497

Sprayberry, D., Impey, C., Irwin, M., \& Bothun, G. 1995 (in preparation).

Thompson, L. A., \& Gregory, S. A. 1993, \aj, 106, 2197

Thompson, L. A. \& Valdes, F.  1987, \apjl, 315, L35 (TV87)

Turner, J. A., Phillipps, S., Davies, J. I., \& Disney, M. J. 1993,
 \mnras, 261, 39

Tyson, J. A. 1986, J. Opt. Soc. Am. A, 3, 2131

Tyson, J. A. 1988, \aj, 96, 1

Tyson, J. A. \& Seitzer, P.  1988, \apj, 335, 552 (TS88)

Ulmer, M. P., Wirth, G. D., \& Kowalski, M. P. 1992, \apj, 397, 430

Ulmer, M. P., Nichol, R. C., Bernstein, G. M., \& Tyson, J. A. 1995 (in
 preparation)

van den Bergh, S., 1992, \aap, 264, 75.

Vader, J. P., \& Sandage, A. 1991, \apjl, 379, L1

Valdes, F. 1989, in Proc. 1st ESO/ST-ECF Data Analysis Workshop, edited by
	P. J. Grosbol, F. Murtagh, \& R. H. Warmels (ESO:  Garching)

White, S.D.M., Briel, U. G., and J. P. Henry, 1993, MNRAS, 261, L8

White, S. D. M., Navarro, J. F., Evrard, A. E., \& Frenk, C. S. 1993,
Nature, 366, 429

Zaritsky, D, Smith, R., Frenk, C., \& White, S. D. M. 1993, \apj, 405, 464

\end{references}
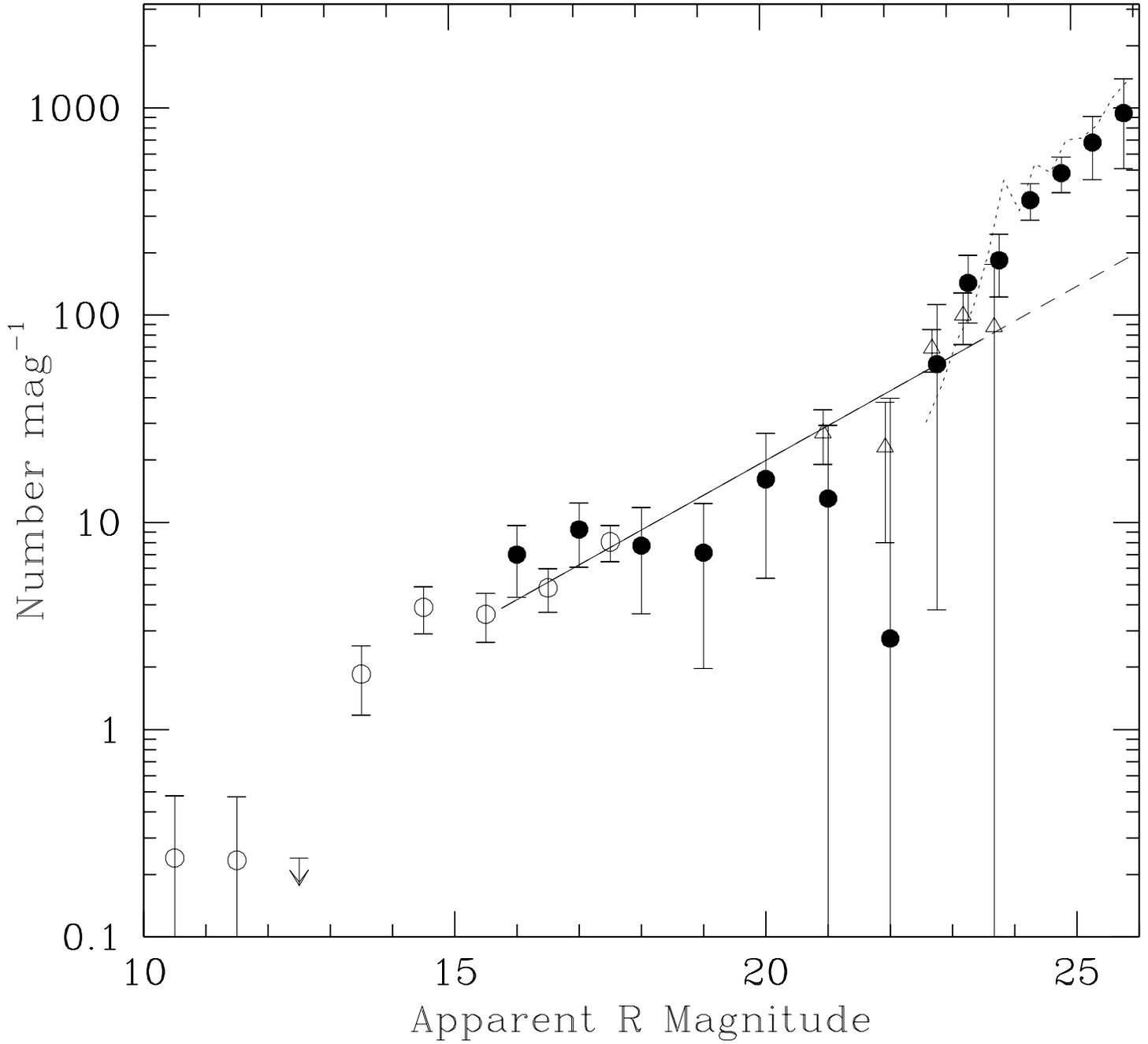
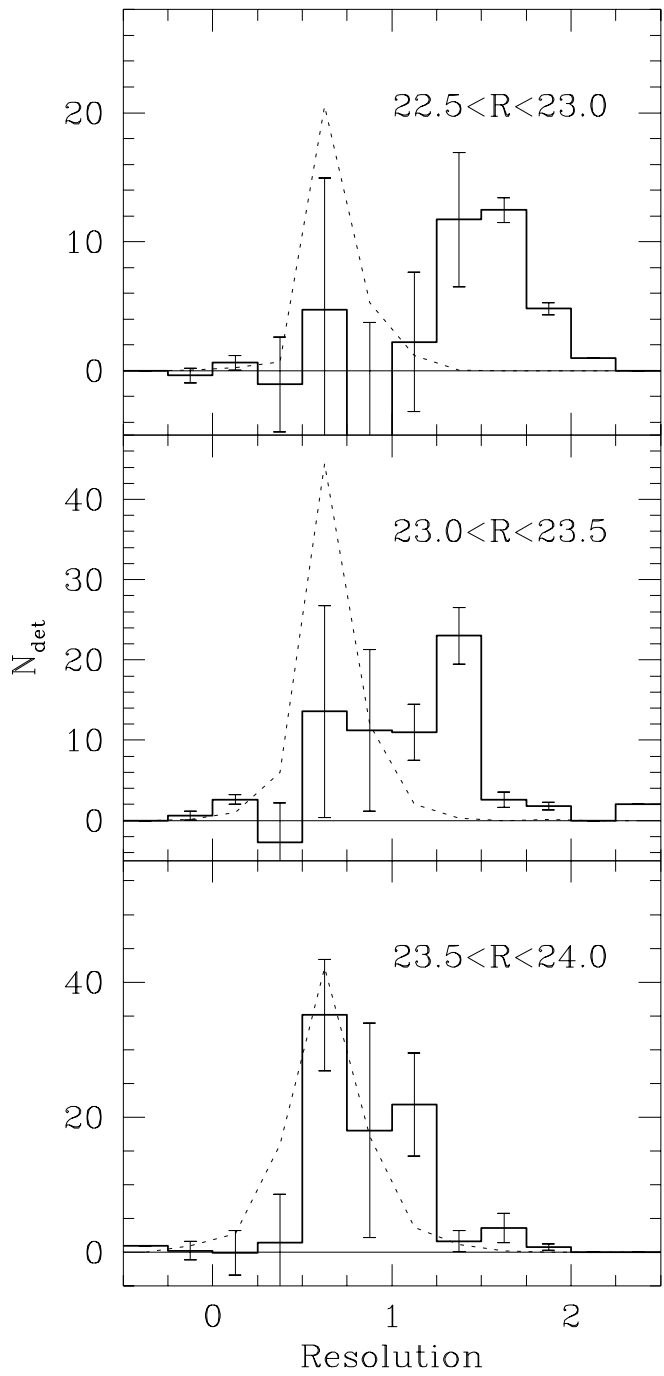
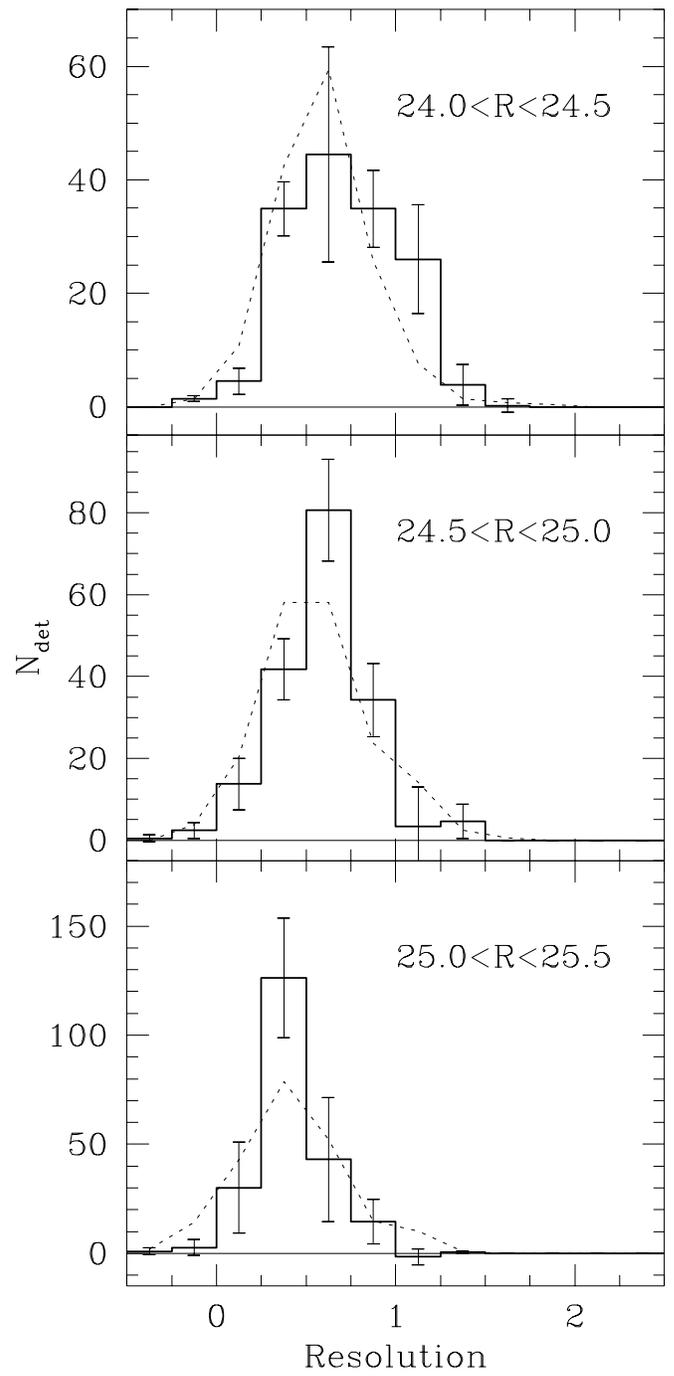
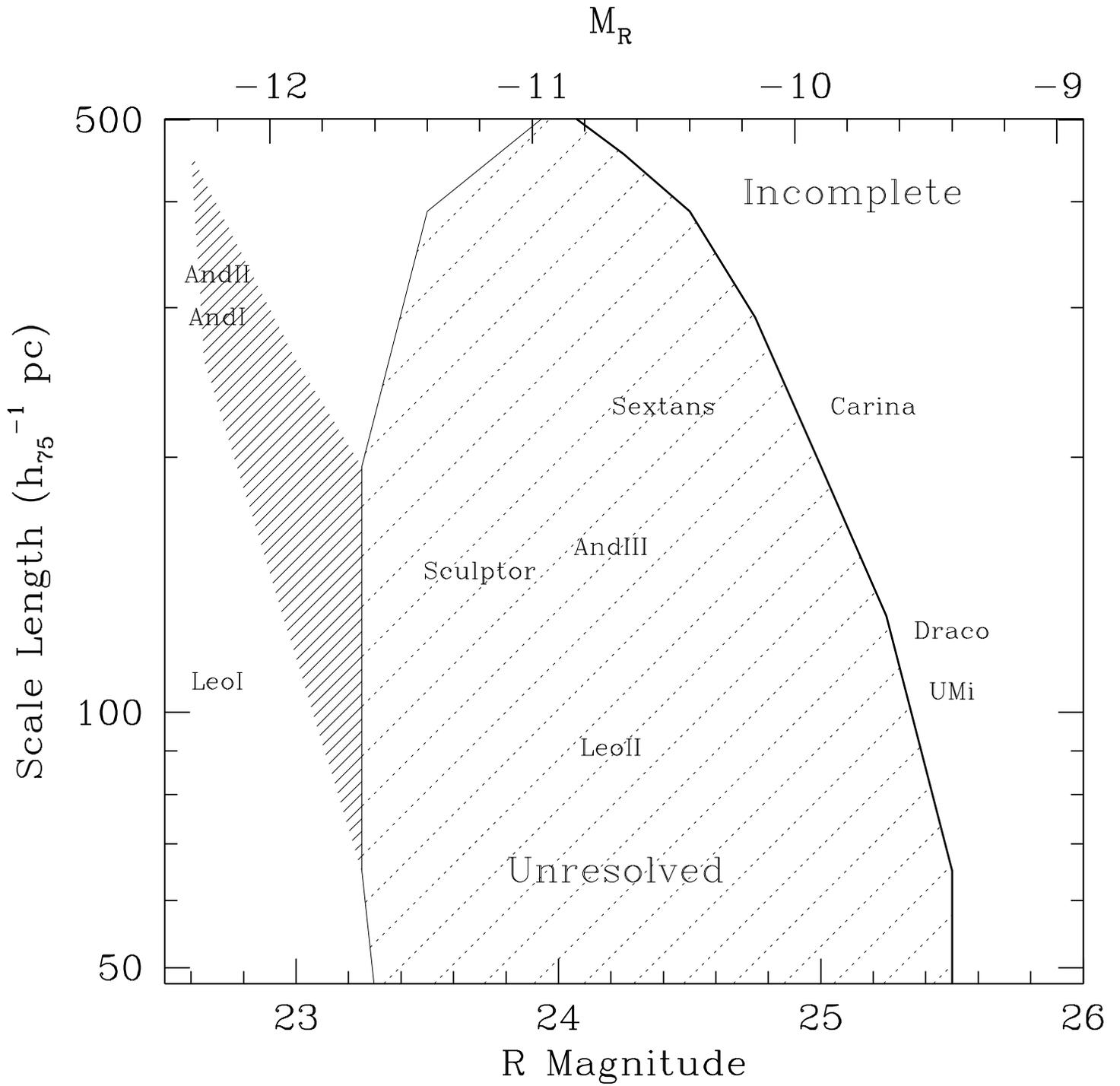
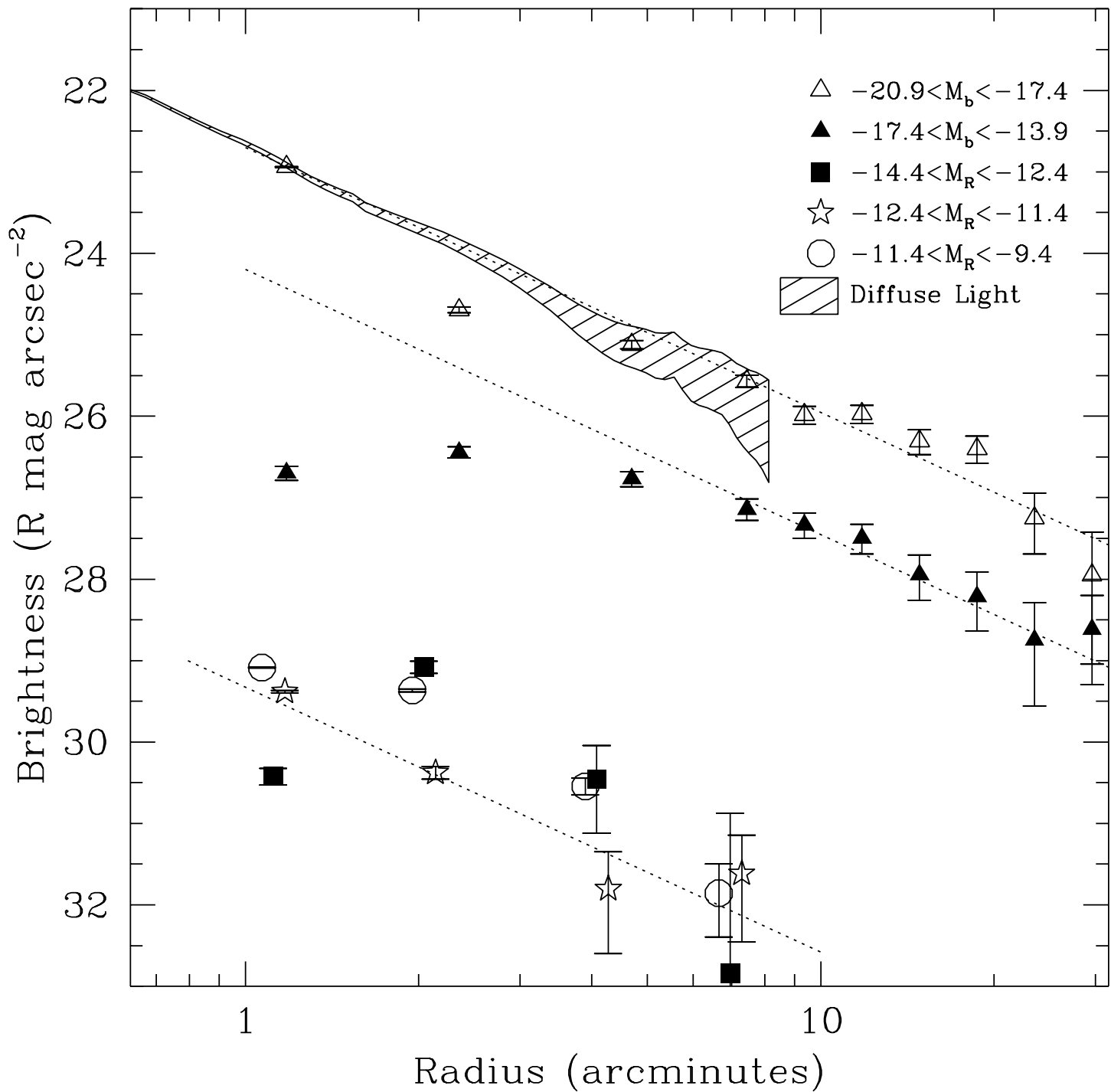
\end{document}